\documentclass[12pt,preprint,preprintnumbers,amsfonts,aps,nofootinbib]{article}
\pdfoutput=1 
\usepackage{amssymb}
\usepackage{amsmath}
\usepackage{epsfig}
\usepackage{epsf}
\usepackage{graphicx}
\usepackage{amsfonts}
\usepackage{amssymb}
\usepackage{cite}
\usepackage{float}
\usepackage{ulem}
\usepackage{hyperref}
\hypersetup{
    colorlinks=true,
    linkcolor=black,
    filecolor=blue,      
    urlcolor=blue,
citecolor=blue}
\usepackage{lipsum}
\usepackage[font=footnotesize]{caption}
\usepackage{subcaption}
\usepackage{soul}
\usepackage[dvipsnames]{xcolor}
\usepackage{todonotes} % for comments
\newcommand\Tucker[1]{\textcolor{green}{\textbf{Tucker:} #1}}

\def\tr{\tilde{r}}
\def\tL{\tilde{\Lambda}}
\def\tE{\tilde{E}}

\makeatletter
\renewcommand\section{\@startsection {section}{1}{\z@}%
                                 {-3.5ex \@plus -1ex \@minus -.2ex}%nn
                                   {2.3ex \@plus.2ex}%
                                   {\normalfont\large\bfseries}}
\renewcommand\subsection{\@startsection{subsection}{2}{\z@}%
                                   {-3.25ex\@plus -1ex \@minus -.2ex}%
                                     {1.5ex \@plus .2ex}%
                                     {\normalfont\bfseries}}
\renewcommand\subsubsection{\@startsection{subsubsection}{3}{\z@}%
                                   {-3.25ex\@plus -1ex \@minus -.2ex}%
                                     {1.5ex \@plus .2ex}%
                                     {\normalfont\itshape}}
\makeatother

\def\beq{\begin{equation}}
\def\eeq{\end{equation}}
\def\be{\begin{equation}}
\def\ee{\end{equation}}
\def\bea{\begin{eqnarray}}
\def\eea{\end{eqnarray}}

\DeclareRobustCommand{\SkipTocEntry}[4]{}

\numberwithin{equation}{section}

\begin{document}

\begin{titlepage}

\setcounter{page}{1} \baselineskip=15.5pt \thispagestyle{empty}

\begin{flushright}
%IPMU-10-xxxx
\end{flushright}
\vfil

\begin{center}

{\Large \bf The classical double copy of  \\ non-singular  black holes}
%{\Large \bf  Regular non-singular black holes and their classical double copy}
\\[0.9cm]
{Damien A. Easson$^\dag$, Cynthia Keeler$^\ast$ and Tucker Manton$^\star$}
\\[0.5 cm]
%\vspace{0 cm}
{\small {\sl  Department of Physics,  
Arizona State University, Tempe, AZ 85287-1504, USA}}\\

\vspace{.3cm}

\end{center}

\vspace{.8cm}

%=====================================================================
%=====================================================================
%=====================================================================
\hrule \vspace{0.3cm}
{\small  \noindent \textbf{Abstract} \\[0.3cm]
\noindent We apply the classical double copy procedure to a class of regular, non-singular black hole solutions. We give several examples, paying particular attention to a string-theory-corrected black hole solution emerging from T-duality. Non-perturbative stringy corrections introduce an ultraviolet (UV) zero-point length cutoff which results in non-singular black hole spacetimes. Apart from the UV regulator, the solution is equivalent to the Bardeen black hole spacetime. We extend this solution to include an asymptotic de Sitter background. All Yang-Mills field theory quantities associated with the double copy are well-behaved and finite for all values of parameters. We present a thorough analysis of the black hole horizon structure, additionally uncovering a simple yet new connection between horizons on the gravity side and electric fields on the gauge theory side of the double copy.

%Hawking radiation from the black hole reaches a maximum and then cools to a stable remnant in de Sitter space.
}\vspace{0.5cm}  \hrule

\vfil

\begin{flushleft}
{\normalsize { \sl \rm \small{$^\dag$~easson@asu.edu \\ $^\ast$~keelerc@asu.edu \\ $^\star$~tucker.manton@asu.edu}}}\\
%\today
%Month dd, yyyy
\end{flushleft}

\end{titlepage}

\newpage
\tableofcontents
\newpage

%=====================================================================
%=====================================================================
%=====================================================================

%=====================================================================
%=====================================================================
\section{Introduction}
%=====================================================================
%=====================================================================

In general relativity (GR), the only spherically symmetric vacuum solution is the Schwarzschild solution. The resulting Schwarzschild black hole is plagued with singular curvature invariants at its center, indicating a break down of general relativity, a notion made mathematically rigorous by the seminal work of Hawking, Ellis, Penrose, and others during the 1960's \cite{Penrose:1964wq, Hawking:1966sx, Hawking:1966jv, Ellis:1968vy, Hawking:1969sw} (see \cite{Senovilla:2014gza} for a somewhat recent review).  The singularity has prompted physicists to consider gravitational theories beyond GR. Non-singular, spherically symmetric solutions using an antisymmetric, Hermitian metric were found in \cite{Moffat:1978tr}, for example. Additional successes include considering stringy $\alpha '$ corrections as in \cite{Cano:2018aod}, extending Einstein-Maxwell theory to include coupling to higher order curvature terms \cite{Cano:2020ezi}, and introducing interacting gauge fields in Lovelock gravity \cite{Cisterna:2020rkc}.

Alternatively, supplementing the Einstein-Hilbert Lagrangian with `non-standard' matter actions, such as those of an Abelian gauge field in a non-linear electrodynamic theory (NLED), produces non-singular black hole solutions.\cite{Ghosh:2020ece,AyonBeato:1998ub,AyonBeato,Ma:2015gpa,Singh:2019wpu}. Solutions of this type include the Dymnikova metric \cite{Dymnikova:1992ux} and the Bardeen black hole \cite{bardeen:1968}, both of which are sourced by non-singular magnetic monopoles\footnote{In fact in \cite{AyonBeato:1998ub}, the idea to couple a black hole to a NLED source in Einstein gravity was first introduced, while in \cite{AyonBeato}, the authors showed how the Bardeen solution can be understood as being sourced by such a monopole. We review this in section \ref{monopolegravsourceSection}.}. The monopole charge $g$  acts as a regulating parameter, curing the curvature divergences at the center of the black hole. Although the monopole sources are unusual in that the associated Lagrangians are somewhat ad-hoc, they indeed produce reasonable stress energy tensors satisfying the weak energy condition (WEC) \cite{Mars:1996,Borde:1996df,Balart:2014jia}. 

Another possibility for resolving the black hole singularity is Markov's limiting curvature hypothesis (LCH) which suggests, given the existence of a fundamental ultra-violet (UV) length scale, all curvature invariants remain bounded \cite{markov}. The LCH has been used to construct non-singular black holes in a variety of contexts~\cite{Frolov:1988vj, Frolov:1989pf, Morgan:1990yy, Banks:1992xs, Trodden:1993dm,Easson:2001qf,Grumiller:2002nm,Easson:2002tg, Chamseddine:2016ktu,  Easson:2017pfe,  Yoshida:2018kwy}   and to resolve cosmological singularities as well~\cite{Mukhanov:1991zn,Brandenberger:1993ef,Moessner:1994jm,Brandenberger:1995es,Brandenberger:1998zs,Easson:1999xw,Easson:2003ia,Chamseddine:2016uef,Easson:2006jd}. In addition, non-singular, or regular, black hole metrics  are appealing through the lens of black hole thermodynamics \cite{Man:2013hpa}. Unlike the Schwarzschild black hole, a generic feature of non-singular solutions is a finite, non-zero final evaporation temperature \cite{Carballo-Rubio:2018pmi}. The Hayward model \cite{Hayward:2005gi}, for example, nicely illustrates the attractive physical characteristics of non-singular black hole evaporation processes. The non-singular nature of the black holes has important implications for the evaporation process, remnants, dark matter and the information loss problem, all of which are discussed partly in the above references. 

Another common thread between the non-singular black hole models is the existence of a ``de Sitter core''; the spacetime geometry locally becomes de Sitter space as the radial coordinate $r$ is taken to zero. The effective cosmological constant in the small $r$ region is related to the metric parameter that serves to regulate the curvature singularities. It is natural to expect the regulating parameter to come from some quantum gravity theory, and indeed in \cite{Nicolini:2019irw}, it was shown that a non-perturbative string theory correction produces an ultra-violet cutoff parameter $\ell$, known in the literature as a zero-point length, which appears in the denominator of the gravitational potential. The parameter $\ell$ plays the precise role of the singularity-regulating parameter in the metric. The resulting spacetime is non-singular and equivalent in form to the Bardeen black hole with $\ell$ in place of the monopole charge $g$. Their derivation can be understood as induced by T-duality, and hence, we refer to their solution as the T-duality black hole.

The double copy, first presented in \cite{Bern1}, provides a simple way to obtain graviton scattering amplitudes in terms of simpler gauge theory amplitudes (see \cite{Bern2} for a comprehensive review on the subject). Qualitatively, the double copy entails first organizing a diagrammatic expansion in (super) Yang-Mills theory such that the overall amplitude manifestly exhibits a particular color-kinematic duality, the BCJ duality, then replacing the non-Abelian gauge theory color factors with `kinematic numerators' (momenta invariants and polarizations). The resulting gravitational amplitude will match both the brute force calculation using the expansion of the Ricci scalar at the Lagrangian level in linearized gravity as well as that obtained using the string theory KLT relations (see respectively sections 1.1 and 2.3.1 of \cite{Bern2}). The double copy is inherently perturbative at the amplitude level; however, in \cite{MonteiroMainCDC}, it was shown that a double copy relationship exists between exact solutions in gravity and exact solutions in gauge theory. This part of the story is referred to as the classical double copy, and will be the focus of our section \ref{NSBHdoublecopy}. 

The classical double copy has been studied in wide variety of settings, \cite{LunaTypeD,LunaTaubNUT,KimPointCharge,LeeDoubleFT1,LeeDoubleFT2,CarrilloGonzalez3Dcdc, LunaSelfDual,TroddenMaxSym,LunaCurvedBckgrnd,Alfonsi:2020lub,Adamo:2017nia,Ilderton:2018lsf,Gurses:2018ckx,Bah:2019sda,Andrzejewski:2019hub,Goldberger:2019xef,Bahjat-Abbas:2020cyb}, including in the context of non-singular, static spherically symmetric black holes \cite{WiseSphericallySymCDC}. Overall findings are generally consistent with what one would expect intuitively (the textbook examples being a point charge mapping to Schwarzschild and Kerr being associated with electric and magnetic fields, the latter depending on the rotation parameter). It has also been extremely fruitful in the realm of gravitational wave physics from black hole collisions \cite{Goldberger:2017frp,Luna:2017dtq,Goldberger:2017vcg,Shen:2018ebu,Cheung:2018wkq,Kosower:2018adc,Bern:2019nnu,Bern:2019crd}. Recent explorations of the classical double copy include its interplay in the fluid-gravity duality \cite{Keeler:2020rcv}, classical backreaction \cite{Adamo:2020qru}, as well as in the context of Born-Infeld theory \cite{Pasarin:2020qoa}.

In this work, we explore non-singular black hole solutions in the framework of the classical double copy. We focus on the T-duality black hole of \cite{Nicolini:2019irw}; however we study a slightly more general metric with de Sitter asymptotics. The metric, which we call the T-dual-de Sitter black hole, is structurally identical to the Bardeen-de Sitter metric studied in \cite{Fernando:2016ksb}, although arguably is derived from a more well-motivated origin%
\footnote{Note the case of de Sitter asymptotics no longer obeys T-duality.},
 since the black hole regulator comes from the UV corrected massless scalar propagator. Associated with the metric $g_{\mu\nu}$ is an Abelian gauge field $A_\mu$ that is referred to as the single copy. We will show that for the T-dual-de Sitter black hole spacetime, all gauge field quantities are non-singular at the origin:
\begin{equation}\label{finitegaugequantities}
\lim_{r\rightarrow 0}A_\mu=0, \ \ \ \ \ \ \lim_{r\rightarrow 0}F_{\mu\nu}=0, \ \ \ \ \ \ \lim_{r\rightarrow 0}\partial_\nu F^{\mu\nu}=\text{constant},
\end{equation}
where $F_{\mu\nu}=\partial_\mu A_\nu-\partial_\nu A_\mu$ is the usual field strength tensor. We will see that verifying (\ref{finitegaugequantities}) is essentially a trivial task, although not necessarily obvious a priori.

Moreover, by investigating the T-dual-de Sitter black hole we find illuminating relationships between the electric field associated with the single copy gauge theory and horizons on the gravitational side. The horizon structure can be understood in terms of two parameters in the metric, and the spacetime can exhibit between zero and six horizons. In section \ref{sec:nsbh}, we present a complete analysis of the (positive mass, positive cosmological constant) parameter space for the T-dual-de Sitter black hole spacetime and find a natural classification scheme that breaks the metric's behavior into 12 total cases (see Fig. \ref{alphaVsLambda}). Following a comprehensive treatment of the causal structure, energy conditions, and circular orbits of the black hole, we show how the single copy electric fields exhibit significantly different behaviors that can be traced back to the horizon structure of the black hole. 
%To the best of our knowledge, these connections have yet to be discussed in the double copy literature. 

This paper is organized as follows. Section \ref{sec:nsbh} is devoted to the features of the T-dual-de Sitter black hole spacetime most relevant to our treatment of the associated single copy gauge theory: we analyze the horizon structure in \ref{HorizonStructureSurfaceGrav}, present the conformal diagrams in \ref{PenroseDiagrams}, work through the energy conditions in \ref{EnergyConditions} including a review of the NLED monopole source for the Bardeen solution, then treat massless and massive circular orbits in \ref{Orbits}. In section \ref{NSBHdoublecopy}, we start with a short review of the classical double copy before computing the single copy gauge field, field strength, and charge density. We also include a discussion of the interplay between the sources on the gravity and gauge theory sides, the Komar energy, along with a few comments about the Hayward and Dymnikova solutions towards the end of section \ref{KSscSection}. In section \ref{VanEfieldsAndHorizons}, we illustrate our new findings relating the single copy electric fields to the horizon structure before adding a few last statements regarding electric and gravitational forces in section \ref{forcessection}. We conclude in section \ref{conclusion}.

%=====================================================================
%=====================================================================
\section{Non-singular black hole solution}\label{sec:nsbh}
%=====================================================================
%=====================================================================

\iffalse
We are inspired by the T-duality black hole solution found in \cite{Nicolini:2019irw}; however, we study a more general solution with de Sitter asymptotics\footnote{\Tucker{Scalar field perturbations and a few other properties of a nearly identical metric were studied in \cite{Fernando:2016ksb}}}. Our nonsingular static and spherically symmetric black hole solution takes the form 
\fi

The T-dual-de Sitter black hole metric can be written as 

\be\label{metricg}
ds^2 = g_{tt}(r) dt^2 + g_{rr}(r)dr^2 + r^2 d \Omega^2 \,,
\ee
with $d \Omega^2 = d\theta^2 + \sin^2\!\theta \,d\phi^2$ and $g_{rr}(r)=-\frac{1}{g_{tt}(r)}$, where the metric function $g_{tt}$ has the profile
\be\label{gttmet}
-g_{tt}(r) = 1 - \frac{2 M(r)}{r}  - \frac{\Lambda r^2}{3} \,,
\ee
with 
\be\label{gfunc}
M(r) = \frac{mr^3}{(r^2 + \ell^2)^{3/2}}
\,.
\ee
In the above, $m$ is the (black hole) mass,  $\ell$ is a new parameter with units of length (associated with the fundamental zero-point length or string scale $\ell \sim \sqrt{\alpha^\prime} $), 
$\Lambda$ is the cosmological constant setting the scale of the de-Sitter space, and our coordinates span the ranges $t \in (-\infty, \, \infty)$, $r \in (-\infty, \, \infty)$, $\theta \in [ 0, \, \pi ]$ and $\phi \in (-\pi,\, \pi]$. Depending on model parameters, the above metric smoothly interpolates between de-Sitter space as $r \rightarrow \pm \infty$, ordinary Schwarzschild for mid-ranges of the radial coordinate, and an ``interior" spacetime, and is everywhere non-singular even as $r \rightarrow 0$. The lack of singularities is most easily seen by examination of the Kretschmann scalar  constructed from the Riemann tensor $R_{\mu \nu \rho \sigma},$ which is everywhere finite:
\bea
 \mathcal{K}&= &R_{\mu \nu \rho \sigma}R^{\mu \nu \rho \sigma} \nonumber \\
&=&
\frac{12 m^2 \left(8 \ell^8-4 \ell^6 r^2+47 \ell^4 r^4-12 \ell^2 r^6+4 r^8\right)}{\left(r^2 +\ell^2 \right)^7}+ \frac{8 \Lambda  m \left(4 \ell^4-\ell^2 r^2\right)}{\left(r^2 +\ell^2\right)^{7/2}}+\frac{8 \Lambda ^2}{3} \,.
\eea

By expanding the metric around $r=0,$ we find the static de Sitter patch
\begin{equation}\label{deSitterCore}
-g_{tt}(r\approx 0)=1-\frac{r^2}{l^2}+O(r^4),
\end{equation}
where $l$ is the interior length scale
\begin{equation}\label{interiorscale}
\frac{1}{l^2}=\frac{\Lambda}{3}+\frac{2m}{\ell^3}.
\end{equation}
The relation (\ref{interiorscale}) suggests some particularly intriguing features for the exotic case of $m<0.$ Depending on the magnitude of the mass, the interior length scale can become infinite if $m=-\frac{\ell^3\Lambda}{6}$, making the interior region locally flat. Alternatively, if $m<-\frac{\ell^3\Lambda}{6},$ the interior region is locally AdS. These features would also arise for $m>0$ and $\Lambda<0$. In this work, we restrict our attention to $m>0$ and $\Lambda>0;$ we plan to explore these other cases in future work.

 Far from the black hole, we find the asymptotic behavior
\begin{equation}
-g_{tt}(r)=1-\frac{2m}{r}+\frac{3m\ell^2}{r^3}-\frac{\Lambda r^2}{3}+O(r^{\text{-}5}).
\end{equation}
This form differs from the standard Reissner-Nordstrom-de Sitter black hole sourced by an electric point charge $q$, since here we have a $r^{-3}$ contribution rather than the $q^2/r^2$ term in $-g_{tt}(r)$. We will return to this point in section \ref{monopolegravsourceSection}.

%=====================================================================
%=====================================================================
\subsection{Horizon structure and surface gravity}\label{HorizonStructureSurfaceGrav}
 %=====================================================================
 %=====================================================================

We plot the metric function (\ref{gttmet}) for all cases in Figures \ref{classMplot}, \ref{classAplot}, and \ref{classesBCplot}. For a general solution there may be an interior horizon, exterior horizon and cosmological horizon, in both the $r>0$ and $r<0$ regions. We denote the location of the three horizons as $r_+,$ $r_-$, and $r_c$, respectively.

 The timelike Killing vector 
 \be\label{tlkv}
 K^\mu = (\partial_t)^\mu = (1,\, 0,\, 0,\, 0) \,
 \ee
is null at each horizon, and has norm
 \be
 K^\mu K_\mu = g_{\mu \nu}  K^\mu K^\nu =  -\left(1 -\frac{2 m r^2}{\left(r^2 +\ell^2 \right)^{3/2}}-\frac{\Lambda  r^2}{3} \right)
 \,,
 \ee
 yielding surface gravities $\kappa_\dagger$, given by:
 \be
 \nabla_\sigma (- K^\mu K_\mu) =   \nabla_\sigma \left(1 -\frac{2 m r^2}{\left(r^2 +\ell^2 \right)^{3/2}}-\frac{\Lambda  r^2}{3} \right) = 2 \kappa_\dagger K_\sigma \,,
 \ee
where $\dagger = \{c \,, + \,, -\}$ for de Sitter, outer and inner horizons, respectively. In the $\{t,r,\theta,\phi\}$ coordinates, since the timelike Killing vector (\ref{tlkv}) vanishes at the horizons and $K^\mu K_\mu\propto g_{tt},$ the zeros of (\ref{gttmet}) provide the horizon locations.

It is useful to write the metric function using the dimensionless coordinate $\tr=r/\ell$ and define the parameters
\begin{equation}\label{dimless}
\alpha=\frac{2m}{\ell}, \ \ \ \ \ \ \tL=\Lambda \ell^2,
\end{equation}
so that 
\begin{equation}\label{dimlessMet}
-g_{tt}=1-\frac{\alpha\tr^2}{(1+\tr^2)^{3/2}}-\frac{\tL}{3}\tr^2.
\end{equation}
The different horizon structures can be understood in terms of critical values of $\alpha$ and $\tL,$ as we will see below. We present the full $\{\alpha,\tL\}$ parameter space in Fig. \ref{alphaVsLambda}.

%For the figure reference to work, it seems one needs to have a space or two between \begin{figure} and the body of the text. Huh.

\iffalse
\begin{figure}[H]
\centering
\begin{subfigure}{.5\textwidth}
  \centering
  \includegraphics[scale=.37]{gttAssMink.png}
\end{subfigure}%
\begin{subfigure}{.5\textwidth}
  \centering
  \includegraphics[scale=.37]{gttAssdeSit.png}
\end{subfigure}
\caption{\textbf{Left panel:} the metric function (\ref{gttmet}) for the asymptotically Minkowski non-singular solutions $(\Lambda=0$), with the four horizon black hole in orange, the critical two horizon case in blue, and the horizonless wormhole in green. The Schwarzschild black hole is plotted in dashed black. \textbf{Right panel:} (\ref{gttmet}) for asymptotically de Sitter solutions, with the six horizon black hole in orange, critical four horizon case in blue, and the two cosmological horizon wormhole in green. The red curve is the critical four horizon black hole (Nariai spacetime), and Schwarzschild-de Sitter is shown in dashed black. A fifth non-singular black hole, the extremal two horizon ``ultra-critical" case is not shown here but instead in Fig. \ref{UCgttplot}.}
\label{metricplot.png}
\end{figure}
\fi

 %=====================================================================
 %=====================================================================
 %=====================================================================
 %=====================================================================
\subsubsection{Asymptotically Minkowski solutions}
 %=====================================================================
 %=====================================================================
  %=====================================================================
 %=====================================================================

 \begin{figure}[H]
 \centerline{\includegraphics[scale=.45]{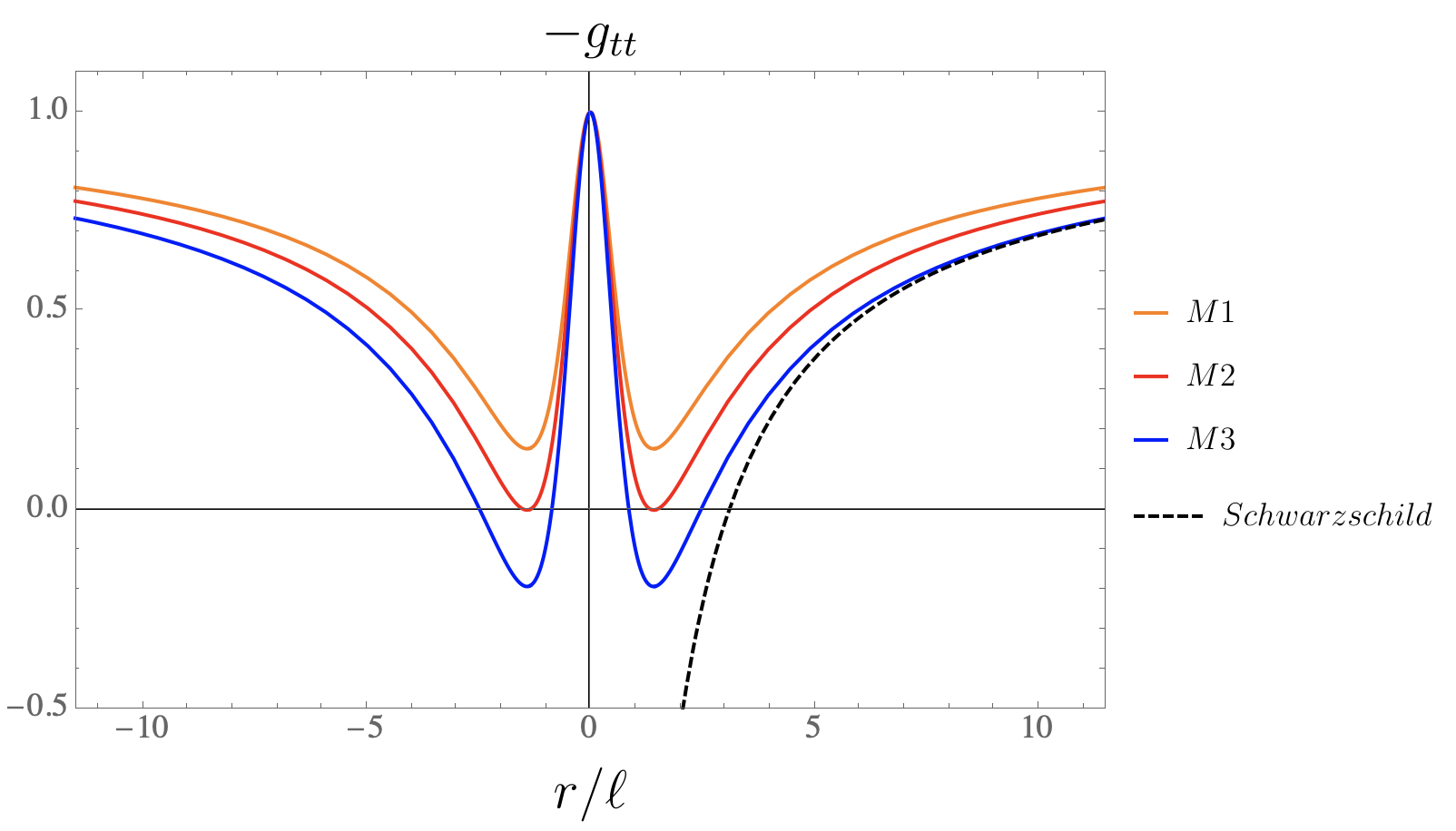}}
\caption{Plot of metric function $-g_{tt}$ for the asymptotically Minkowski spacetimes. The horizonless wormhole is shown in orange, the extremal black hole with merged horizons at $\pm r_+=\pm r_-=\pm r_\star = \pm\sqrt{2}\ell$ in red, and the standard four horizon non-singular black hole in blue. }\label{classMplot}
\end{figure}

When $\Lambda=0,$ the spacetime is asymptotically flat and has the metric profile 
\begin{equation}
-g_{tt}=1-\frac{\alpha \tr^2}{(1+\tr^2)^{3/2}}.
\end{equation}
This solution may have zero, two, or four horizons depending on $\alpha$. The analysis is straightforward: a critical point for $\alpha$ occurs when $-g_{tt}=-\partial_r g_{tt}=0$, corresponding to the value $\alpha_{\text{crit}}=\frac{3\sqrt{3}}{2}.$ At $\alpha_{\text{crit}},$ the inner and outer horizons merge (in both the $r>0$ and $r<0$ regions) at $r=\pm\sqrt{2}\ell$. Consequently, we obtain the following statements for the general horizon structure, which we refer to as the cases $M1,M2$ and $M3$:
 \begin{equation}\label{eqal}
\begin{split}
    M1: \ \ \ \ \ \alpha&<\frac{3\sqrt{3}}{2} \ \ \ \Rightarrow \ \ \ \text{no horizons,} \\
  M2: \ \ \ \ \   \alpha&=\frac{3\sqrt{3}}{2} \ \ \ \Rightarrow \ \ \ \text{two horizons,} \\
 M3: \ \ \ \ \    \alpha&>\frac{3\sqrt{3}}{2} \ \ \ \Rightarrow \ \ \ \text{four horizons}.
\end{split}
 \end{equation}

In the $\{\alpha,\tL\}$ parameter space in Fig. \ref{alphaVsLambda}, the above three cases all lie on the vertical ($\tL=0$) axis, with $M1$ in dashed blue, $M3$ in solid blue, and the $M2$ point at $(\alpha,\tL)=(3\sqrt{3}/2,0)$ labeled with a solid blue star. Fig. \ref{classMplot} shows these three separate cases contrasted to the usual Schwarzschild solution.\footnote{By reinserting $G$ and $c$ so that $\alpha = \frac{2Gm}{\ell c^2}$, we can approximate regime in which two or no horizons would be possible for a physical black hole is $\frac{m}{\ell}\sim \frac{3\sqrt{3}}{4}\frac{c^2}{G}\sim 1.75\times 10^{27} \ \text{kg/m}.$ If $\ell\sim O(\ell_{\text{Planck}}),$ then the black hole mass would be around 30 $\mu$g, roughly the order of the Planck mass. Any macroscopic black hole today described by the model under consideration would have two horizons in the $r>0$ region, although the inner horizon is generally located a few multiples of $\ell$ from the origin $r=0.$ These estimates are consistent with the findings of \cite{Nicolini:2019irw,Fernando:2016ksb}.} In the $M3$ case, we have that $-g_{tt}<0$ between the inner and outer horizons. In this region, the $r$ coordinate switches roles with $t$ and timelike (spacelike) curves become spacelike (timelike) curves.

We have two black hole surface gravities given by
\be\label{surfacek}
\kappa_\pm = \frac{1}{2}  \partial_r \left(1 -\frac{2 m r^2}{\left(r^2 +\ell^2 \right)^{3/2}} \right) \Bigg\rvert_{r_\pm},
\ee
 where the above is evaluated at the either the inner or outer horizons $r_\pm$. In the case where the inner and outer horizons merge to $\pm r_+=\pm r_-\equiv\pm r_{\star},$ it follows that $\partial_r g_{tt}(\pm r_\star)=0$, signaling a vanishing surface gravity. In GR, surface gravity is proportional to the black hole temperature, therefore extremal solutions of this type are often referred to as cold black holes.

  For a given black hole mass $m$, the horizons $r_\pm$ are given by the expression:
 \be
 m = \frac{\left(\ell^2+r_\pm^2\right)^{3/2}}{2 r_\pm^2} \,,
 \ee
 which has global minimum corresponding to the extremal values $r_\star= \sqrt{2} \ell$,  $m_\star = 3\sqrt{3}\ell /4$. In the limit as $m \rightarrow \infty$, the outer horizon $r_+ \rightarrow 2m$ and the inner horizon $r_-  \rightarrow \ell$.

  %=====================================================================
 %=====================================================================
 %=====================================================================
 %=====================================================================
\subsubsection{Asymptotically de Sitter solutions}\label{AssDeSitGravSection}
 %=====================================================================
 %=====================================================================
  %=====================================================================
 %=====================================================================
 
With a nonzero (positive\footnote{We focus on $\Lambda>0$ and $m>0$. We note, however, that unlike negative mass Schwarzschild black holes, negative mass solutions with (\ref{gfunc}) cannot be ruled out simply on the bases of a cosmic censorship conjecture, as horizonless solutions exist without the objectionable singularities. Such solutions violate several if not all of the canonical energy conditions.}) cosmological constant, the metric (\ref{gttmet}) has a very rich horizon structure. It is natural to divide the various cases in to three main classes, which we refer to as classes $A$, $B$, and $C$. The relation between $\alpha$ and $\tL$ that distinguishes between the three classes can be obtained from finding the $r$-values for the extrema of $-g_{tt}$ by computing the zeros of $-\partial_r g_{tt}$ such that $-\partial_r g_{tt}(r_{\text{ex}})=0$, asking whether $-g_{tt}$ is positive, negative, or zero at $r_{\text{ex}}$, and finally answering if $-g_{tt}$ is linear, quadratic, or cubic near $r_{\text{ex}}$. We find that the line
\begin{equation}
\alpha = \frac{5^{5/2}}{3}\tL
\end{equation}
in the $\{ \alpha,\tL\}$ parameter space serves to differentiate between $-g_{tt}$ having either local extrema (class $A$), inflection points (class $B$), or the sole global maxima and no other extrema (class $C$).

 %=====================================================================
 %=====================================================================
\paragraph{Class \textit{\textbf{A}}: four local extrema}
 %=====================================================================
 %=====================================================================

 \begin{figure}[H]
 \centerline{\includegraphics[scale=.45]{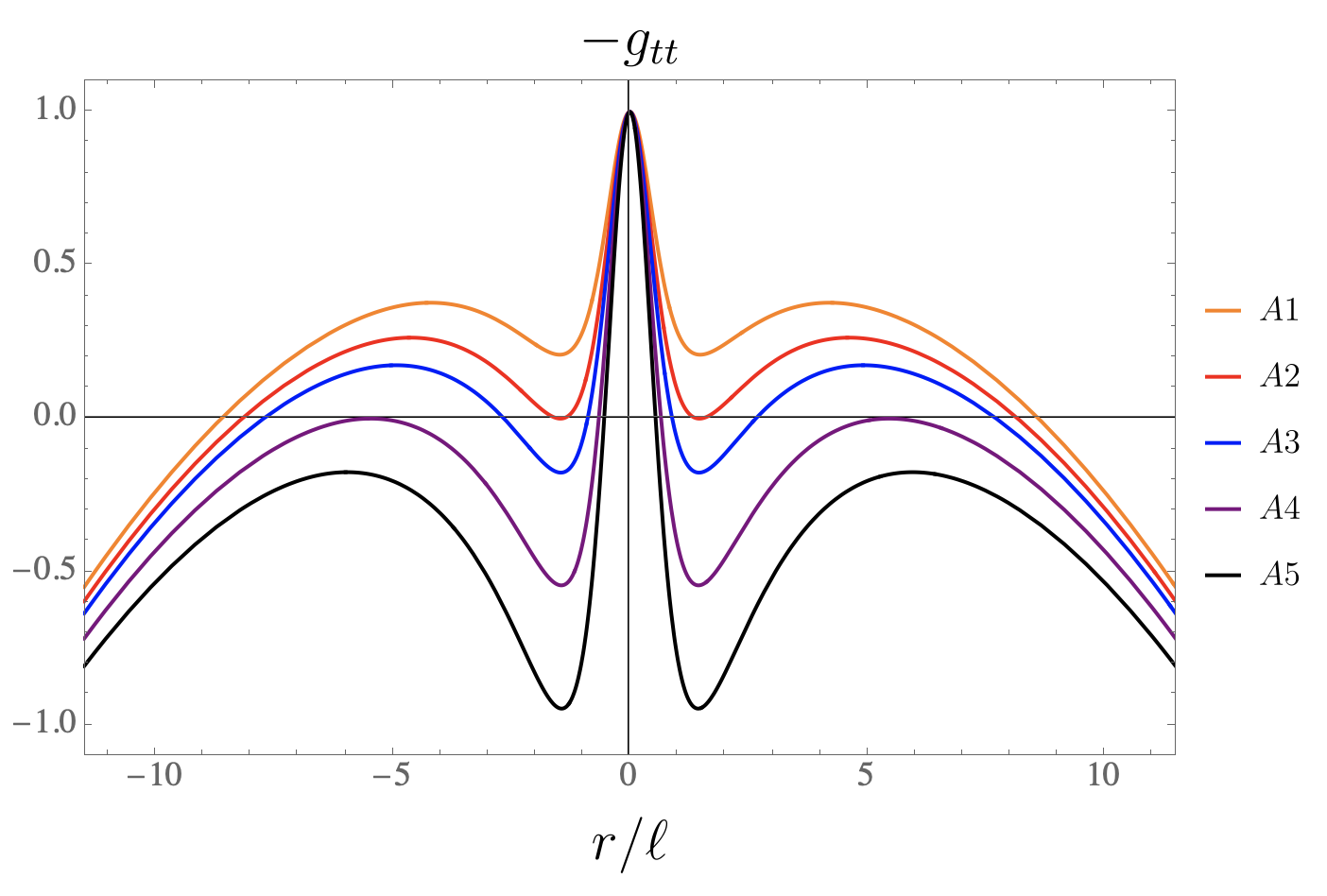}}
\caption{Plot of metric function $-g_{tt}$ for the class A asymptotically de Sitter spacetimes. }\label{classAplot}
\end{figure}

There are five different cases that exhibit four local extrema and one global maximum. These cases may have two, four, or six horizons. We denote the radius where $-g_{tt}$ has a local minimum as $\tr_{\text{min}}$ and the radius where $-g_{tt}$ has a local maximum as $\tr_{\text{max}}$. Since $-g_{tt}$ is symmetric in $r\rightarrow -r,$ there will always be a local extrema at $\pm \tr_{\text{max}}$ and $\pm \tr_{\text{min}}$. The global maximum always occurs at $-g_{tt}(\tr=0)=1.$ For all five cases, we have that 
\begin{equation}
\alpha > \frac{5^{5/2}}{3}\tL.
\end{equation}

 %=====================================================================
 %=====================================================================
\subparagraph{\textit{\textbf{A1}}: Wormhole with cosmological horizons}
 %=====================================================================
 %=====================================================================

If $-g_{tt}(\tr_{\text{min}})>0,$ then the spacetime exhibits no black hole horizons and two cosmological horizons at radii $\pm \tr_c$. An example of such a spacetime is plotted in orange in Fig. \ref{classAplot}. In the parameter space of $\{\alpha,\tL\}$, the $A1$ subclass is bounded by the $\tL=0$ axis, the $\alpha=\frac{5^{5/2}}{3}\tL$ line, and above by a curve $\alpha_{A2}(\tL)$ we will describe in the next subsection. 

 %=====================================================================
 %=====================================================================
\subparagraph{\textit{\textbf{A2}}: Merged black hole horizons}
 %=====================================================================
 %=====================================================================
 
The inner and outer black hole horizons merge when $-g_{tt}$ as in (\ref{dimlessMet}) satisfies
\begin{equation}
-g_{tt}(\tr_{\star})=-\partial_r g_{tt}(\tr_{\star})=0, \ \ \ \ \ -\partial^2_r g_{tt}(\tr_{\star})>0,
\end{equation}
where $\tr_{\star}=\tr_{\text{min}}=\tr_{\text{max}}$ is the location of the merged horizons,
\begin{equation}
\tr_{\star}^2 = \frac{1-\sqrt{1-8\tL}}{2\tL}.
\end{equation}
$-g_{tt}$ is quadratic near $\tr=\pm\tr_\star$. The associated critical value for $\alpha$ is given by 
\begin{equation}\label{alphaA2}
\alpha_{A2}(\tL)=\frac{2\tL}{3}\frac{(\tr_{\star}^2+1)^{5/2}}{\tr^2_{\star}-2},
\end{equation}
which is shown in purple in Fig. \ref{alphaVsLambda}. Spacetimes of this type are also referred to as cold black holes, as $\partial_r g_{tt}\propto T_H=0$ at the merged horizon. Since we still have de Sitter asymptotics, there will also be cosmological horizons at $\pm r_c.$ An example of such a spacetime is shown in red in Fig. \ref{classAplot}.

 %=====================================================================
 %=====================================================================
\subparagraph{\textit{\textbf{A3}}: Four black hole and two cosmological horizons}
 %=====================================================================
 %=====================================================================

The metric (\ref{dimlessMet}) can exhibit at most six horizons. This subclass is analogous to $M3$ in the asymptotically flat examples except with cosmological horizons for large $\tr$. The profile function $-g_{tt}$ is linear near all of the six horizons. The $A3$ region in the $\{\alpha,\tL\}$ parameter space is bounded by the $\tL=0$ axis, below by $(\ref{alphaA2})$, and above by a second curve $\alpha_{A4}(\tL)$ that we will describe in the following subsection. We plot an example of such a spacetime in blue in Fig. \ref{classAplot}. In this case, surface gravity can be computed at both the inner and outer black hole horizons, as $-\partial_r g_{tt}\neq 0$ at all of the horizon radii. This class of black holes thus have nonzero temperature.

 %=====================================================================
 %=====================================================================
\subparagraph{\textit{\textbf{A4}}: Merged black hole and cosmological horizons (Nariai spacetime)}
 %=====================================================================
 %=====================================================================
 
 It is possible for the outer black hole horizon to merge with the cosmological horizon. Such solutions are called Nariai spacetimes \cite{Nariai1,Nariai2}, and the metric function satisfies
\begin{equation}
-g_{tt}(\tr_N)=-\partial_r g_{tt}(\tr_N)=0, \ \ \ \ \ -\partial^2_r g_{tt}(\tr_N)<0,
\end{equation}
where now,
\begin{equation}
\tr_N^2=\frac{1+\sqrt{1-8\tL}}{2\tL}
\end{equation}
gives the locations of the merged horizons. $-g_{tt}$ is quadratic near $\pm\tr_N.$ There is still a pair of inner black hole horizons at $\pm\tr_-,$ where $-g_{tt}$ behaves linearly. We therefore can compute nonzero surface gravity at the inner horizon, but not at the merged outer/cosmological horizon.

As for the $A2$ subclass, such spacetimes lie on a curve in the $\{\alpha,\tL\}$ parameter space given by 
\begin{equation}\label{alphaA4}
\alpha_{A4}(\tL)=\frac{2\tL}{3}\frac{(\tr_{N}^2+1)^{5/2}}{\tr^2_{N}-2}.
\end{equation}
This class is shown in orange in Fig. \ref{alphaVsLambda}, while we plot an example of $-g_{tt}$ for such a solution in purple in Fig. \ref{classAplot}. Note that $\lim_{\tL\rightarrow 0}\alpha_{A4}(\tL)\rightarrow \infty,$ as expected; if a cosmological horizon is preserved as $\tL\rightarrow 0$, its location must tend to $\pm\tr_c\rightarrow\pm\infty.$ Keeping the (outer) black hole horizon equal to the cosmological horizon in that limit requires that the black hole mass diverge to infinity, consistent from the intuition built from the Schwarzschild black hole.

 %=====================================================================
 %=====================================================================
\subparagraph{\textit{\textbf{A5}}: Two linear horizons}
 %=====================================================================
 %=====================================================================

Finally, it is possible for $-g_{tt}$ to have all local extrema below the $-g_{tt}=0$ axis. In this case, there is a single pair of horizons, and $-g_{tt}$ is linear about each. The only region in which the spacetime is timelike ($-g_{tt}>0$) is the small-$r$ de Sitter patch, otherwise, $-g_{tt}<0$ for all $r\notin [-r_-,r_-]$. The horizons appear as cosmological horizons to an observer in the small-$r$ region, while they are more similar to the inner black hole horizons to any other observer.

These solutions occupy a significant region of the $\{\alpha,\tL\}$ parameter space. The $A5$ region is bounded below by $\alpha_{A4}(\tL)$ as in (\ref{alphaA4}) and the $\alpha =\frac{5^{5/2}}{3}\tL$ line, but is otherwise unbounded from above. This is easily seen in Fig. \ref{alphaVsLambda}. We plot an example of $-g_{tt}$ for such a spacetime in black in Fig. \ref{classAplot}.

 %=====================================================================
 %=====================================================================
 %=====================================================================
 %=====================================================================
\paragraph{Class \textit{\textbf{B}}: single inflection point}
 %=====================================================================
 %=====================================================================
  %=====================================================================
 %=====================================================================

For $-g_{tt}$ to have an inflection point at some radius $\tr_I,$ we must have that $-\partial_r g_{tt}(\tr_I)=-\partial^2_r g_{tt}(\tr_I)=0.$ Such solutions all lie on the straight line 
\begin{equation}\label{redline}
\alpha = \frac{5^{5/2}}{3}\tL
\end{equation}
in the $\{\alpha,\tL\}$ parameter space. We will see in section \ref{NSBHdoublecopy} that these solutions are unique from the $A$ and $M$ classes when they are considered through the double copy mapping.

  \begin{figure}[H]
 \centerline{\includegraphics[scale=.45]{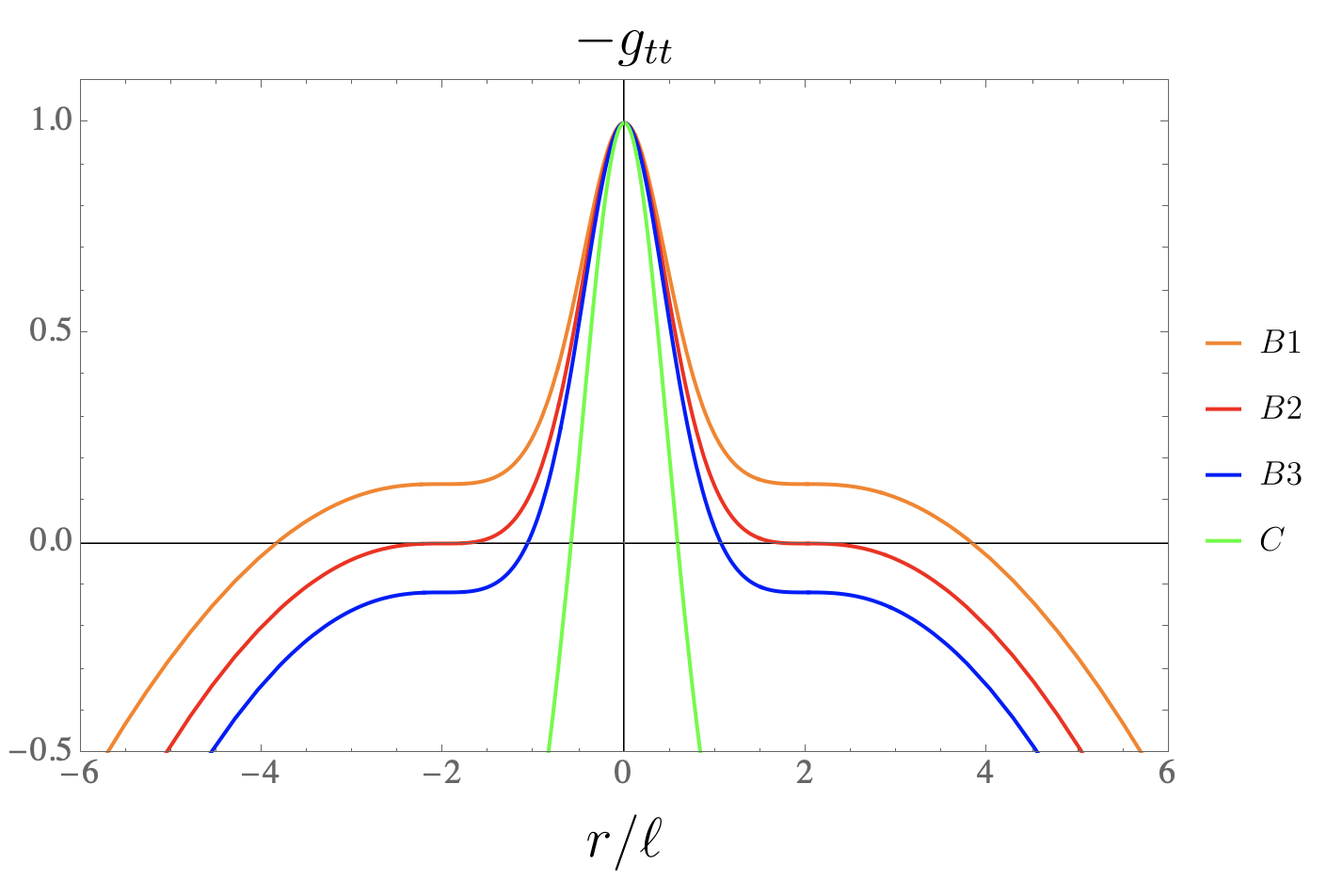}}
\caption{Plot of metric function $-g_{tt}$ for the class $B$ and $C$ asymptotically de Sitter spacetimes. The $B2$ subclass (defined by having parameter values $\{\alpha,\tL\}= \frac{5^{5/2}}{24},\frac{1}{8}\}$) clearly has its horizons at $r/\ell=\pm 2$.}\label{classesBCplot}
\end{figure}

 %=====================================================================
 %=====================================================================
\subparagraph{\textit{\textbf{B1}} and \textit{\textbf{B3}}: linear horizons}
 %=====================================================================
 %=====================================================================
 
Subclasses $B1$ and $B3$ are characterized respectively by $-g_{tt}(\tr_I)>0$ and $-g_{tt}(\tr_I)<0$. Consequently, $-g_{tt}$ will be linear near the horizon location, and a nonzero surface gravity can be computed for both subclasses. Similar to the $A5$ subclass, $-g_{tt}$ has a negative slope near the horizon in the $\tr>0$ region, akin to what we see as either the inner black hole horizon or cosmological horizon in the $A3$ case.

Since the inflection point at $\pm\tr_I$ occurs before the horizon for the $B1$ subclass, it is natural to think of its horizon as being a cosmological horizon for all observers. On the other hand, since the inflection point occurs after the horizon for the $B3$ subclass, it may be more natural to consider its horizon most similar to the inner black hole horizon for observers in the spacelike $(-g_{tt}<0)$ region, and the usual de Sitter core for observers at small $\tr$.

 An example of a $B1$ and $B3$ spacetime is shown in respectively orange and blue in Fig. \ref{classesBCplot}. In the $\{\alpha,\tL\}$ parameter space in Fig. \ref{alphaVsLambda}, the $B1$ subclass lies on the dashed section of the red line given by (\ref{redline}), while the $B3$ subclass lies on the solid section.

 %=====================================================================
 %=====================================================================
\subparagraph{\textit{\textbf{B2}}: cubic horizon}
 %=====================================================================
 %=====================================================================
 
 For $-g_{tt}$ to have a ``cubic" horizon, we have $-g_{tt}(\tr_I)=-\partial_r g_{tt}(\tr_I)=-\partial^2_r g_{tt}(\tr_I)=0.$ This leads to $\tr_I=\pm 2,$ as well as constraining $\alpha$ and $\tL$ to take the values
 \begin{equation}
 \alpha=\frac{5^{5/2}}{24}, \ \ \ \ \ \ \tL=\frac{1}{8},
 \end{equation}
 which is shown as a red star in Fig. \ref{alphaVsLambda}. $-g_{tt}$ has an inflection point at the radius at which $-g_{tt}=0,$ therefore $-g_{tt}$ is approximately cubic near the horizon. It is thus natural to think of this spacetime as having a triply-merged horizon, where the inner, outer, and cosmological horizons all meet at $\pm\tr_-=\pm\tr_+=\pm\tr_c=\pm\tr_I.$ These solutions are sometimes referred to as ultra-cold black holes, since not only is $-\partial_r g_{tt}(\tr_I)=0,$ implying zero temperature, but $-\partial^2_r g_{tt}(\tr_I)=0$ as well. We illustrate an example of such a spacetime in red in Fig. \ref{classesBCplot}.

 %=====================================================================
 %=====================================================================
  %=====================================================================
 %=====================================================================
\paragraph{Class \textit{\textbf{C}}: no local extrema}
 %=====================================================================
 %=====================================================================
  %=====================================================================
 %=====================================================================

The last possibility is for $-g_{tt}$ to be monotonically decreasing from the global maximum of $-g_{tt}(\tr=0)=1$. These solutions occur when the cosmological constant is comparable or larger than the mass of the black hole. Specifically, these solutions occupy the entire
\begin{equation}
\alpha < \frac{5^{5/2}}{24}\tL
\end{equation}
region of the $\{\alpha,\tL\}$ parameter space (see Fig. \ref{alphaVsLambda}). We plot an example of such a solution in green in Fig. \ref{classesBCplot}. It is easy to see that $-g_{tt}$ is approximately linear near the horizons. This spacetime is most similar to pure de Sitter space with `small' cosmological horizons. We will show that the gauge theory associated with this class of solutions through the double copy mapping behaves differently than classes $M$, $A$, or $B$ (see section \ref{EsandRhos}).

 \begin{figure}[H]
 \centerline{\includegraphics[scale=.35]{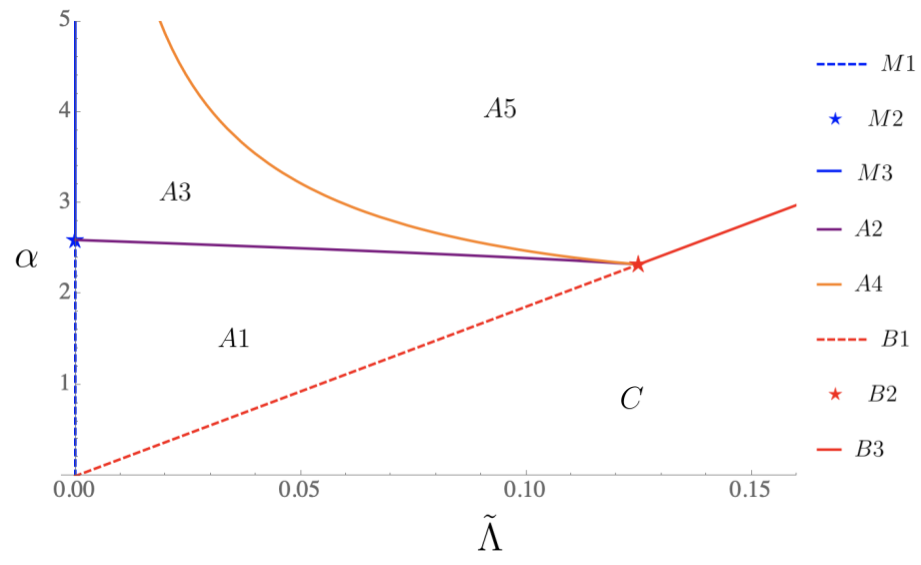}}
\caption{$\{\alpha,\tL\}$ parameter space with all 12 cases. The dashed/solid red line is $\alpha=\frac{5^{5/2}}{3}\tL$, while the purple and orange curves are given respectfully by (\ref{alphaA2}) and (\ref{alphaA4}). The extremal asymptotically flat spacetime ($M2$) is at the point $\{3\sqrt{3}/2,0\}$ and the asymptotically de Sitter `cubic horizon' spacetime ($B2$) is at the point $\{5^{5/2}/24,1/8\}$. We note that $A1-A4$ as well as $B2$ were discussed in \cite{Fernando:2016ksb}, and their findings are consistent with ours for those cases.  } \label{alphaVsLambda}
\end{figure}

%\iffalse
%\begin{figure}
%\centerline{\includegraphics[width=4.75in]{Newgttplot.png}}
%\caption{Plot of the metric function $-g_{tt}$ for various parameter ranges of interest. The %asymptotically Minkowski solutions are the nonsingular black hole with four horizons (orange), the %critical case with two horizons (blue), the horizonless wormhole (green), and Schwarzschild %(dashed green). The asymptotically dS solution with six horizons is shown in red, while the four horizon asymptotically AdS solution is shown in purple. A fifth special case, the  }
%\label{metricplot.png}
%\end{figure}
%\fi
%%%%%%%%%%%%%%%%%%%%%%%%%%%%%%%%%%%%%%%%%%%%%%%%%%%%%%%%%%%%%%%%%%%%%%%%%%%%%%%%%%%%%%%%

%=====================================================================
%=====================================================================
\subsection{Penrose-Carter diagrams}\label{PenroseDiagrams}
%=====================================================================
%=====================================================================

We now construct Penrose-Carter diagrams for the three cases of interest discussed above for the regular black hole solution with Minkowski asymptotics (see Fig.~\ref{classMplot})\footnote{The diagrams can be constructed by defining the usual tortoise function $F(r)=\int^r\frac{dr'}{g_{tt}(r')}$, first passing to coordinates 
$u=t-F,$ $v=t+F,$ then choosing functions $\tilde{u}(u)$ and $\tilde{v}(v)$ that satisfy $g_{tt}(r)\frac{du}{d\tilde{u}}\frac{dv}{d\tilde{v}}>0$ such that the metric is 
$$ds^2=-g_{tt}(r)\frac{du}{d\tilde{u}}\frac{dv}{d\tilde{v}}d\tilde{u}d\tilde{v}+r^2d\Omega^2.$$ See, \textit{e.g.}, \cite{Schindler:2018wbx} for further description.}. The three cases of interest are those parameterized by $\alpha$
 in Eq.~(\ref{eqal}); although, we emphasize the diagrams constructed below are representative of  regular  black hole solutions with de Sitter interior, and not specific to the metric (\ref{gfunc}). As discussed elsewhere in the text, all cases asymptotically approach Schwarzschild at large values of the radial coordinate, and de Sitter spacetime as $r \rightarrow 0$. The most general maximal extension of the regular black hole solution is similar to that of the Reissner-Nordstrom solution but with a non-singular de Sitter interior. In the Penrose-Carter diagrams below, the most general solution describes an infinite number of asymptotically flat regions denoted by $I$ outside of the outer black hole event horizon $r_+$. These regions are connected by intermediate regions: one between the outer
 event horizon and the inner Cauchy horizon $r_-$; denoted by $I \! I$ with  $r_- <r<r_+$, and region $I\! I\! I$ having $0<r<r_-$.  In region $I \! I$, surfaces of constant $r$ are spacelike so that each point in the diagram corresponds to a 2-sphere, $S^2$.  The black hole center at $r=0$ bounds each region $I\! I\! I$ and is non-singular. The light cone structure in the background spacetime is given by null paths with $ds^2=0$ and constant $\theta$ and $\phi$:
\be
\frac{dr}{dt} = \pm \left(1 -\frac{2 m r^2}{\left(r^2 +\ell^2 \right)^{3/2}}-\frac{\Lambda  r^2}{3} \right)
\ee
 
In the diagrams we adopt the traditional notation for spacetime boundaries defined in \cite{Hawking:1973uf}: $\mathcal{J}^-$, $\mathcal{J}^+$ identify past and future null infinity. $i^-$, $i^+$, identify past and future timelike infinity. $i^0$ is spatial infinity while $p$ is used to identify exceptional points at infinity. Without detracting from  our desired analysis, we take $\Lambda=0$ in the diagrams.

 \begin{figure}\label{gttreg.png}
\centerline{\includegraphics[width=4.0in]{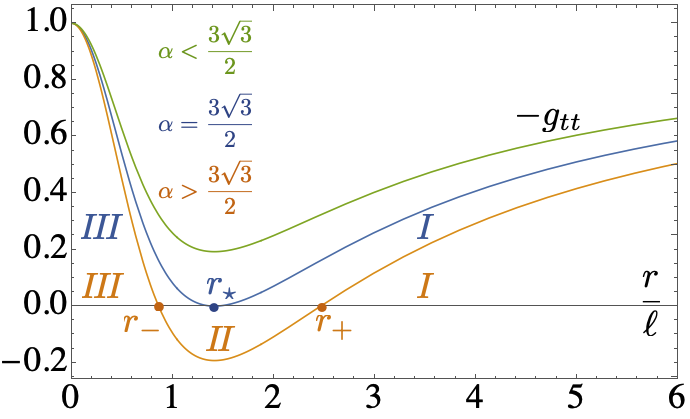}}
\caption{Plot of metric function $-g_{tt}$ versus rescaled radial coordinate $r/\ell$. Shown in the plot are the outer horizon $r_+$, inner horizon $r_-$ and critical horizon $r_\star$. Regions for the three cases of interested are: asymptotically flat region $I$, between horizon region $I \! I$ ~($r_- <r<r_+$), and region~$I\! I\! I$~( $0<r<r_-$).
}
\end{figure}

\break

 %=====================================================================
 %=====================================================================
 \subsubsection{Horizonless wormhole}
 %=====================================================================
 %=====================================================================
 
\begin{figure}[H]
\centerline{\includegraphics[width=3.0in]{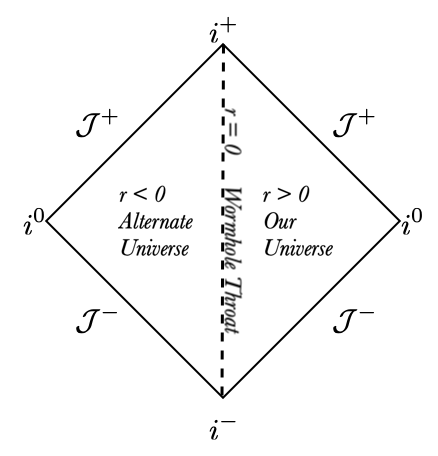}}
\caption{Penrose-Carter diagram for a standard traversable wormhole.
}\label{wormPenrose}
\end{figure}

For the case that   $\alpha <\frac{3\sqrt{3}}{2}$ our solution depicts a two-way traversable wormhole with no horizon. Observers in our universe correspond to observers having $r>0$. From their point of view 
observers living at negative values of the radial coordinate live in an alternate universe.  The Penrose-Carter diagram for this case is depicted in Fig. \ref{wormPenrose}.

\break 
%=====================================================================
%=====================================================================
\subsubsection{Wormhole with single horizon}
%=====================================================================
%=====================================================================

\begin{figure}[H]
\centerline{\includegraphics[width=4.25in]{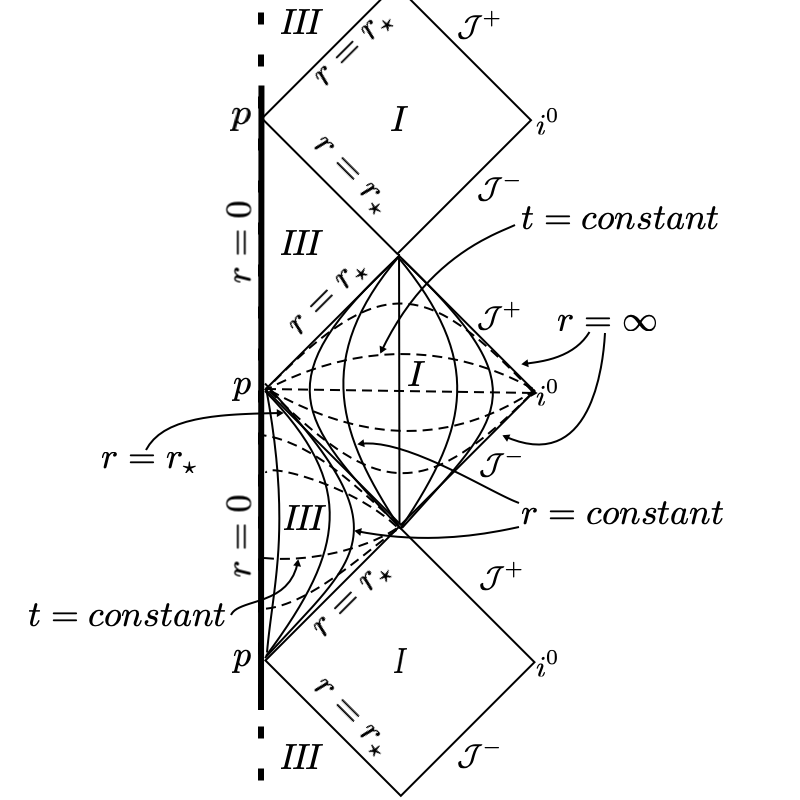}}
\caption{Penrose-Carter diagram for the maximally extended spacetime for regular center black hole (one-way wormhole) when outer and inner horizons coincide.  
}\label{wormcrit}
\end{figure}

For the case that $\alpha = \frac{3\sqrt{3}}{2}$, the solution depicts a one-way wormhole with null throat located at the critical value $r_\star= \sqrt{2} \ell$.  The outer event horizon and inner Cauchy horizon join to form a single horizon at  $r = r_\star$.  There are an infinite set of region $I$s connected by region $I\!I\!I$s. The exceptional points  at infinity $p$ are not actually part of the de Sitter region at $r=0$. The Penrose-Carter diagram for this situation is depicted in Fig. \ref{wormcrit}.

\break 
%=====================================================================
%=====================================================================
\subsubsection{Two horizon black hole}
%=====================================================================
%=====================================================================

\begin{figure}[H]
\centerline{\includegraphics[width=3.75in]{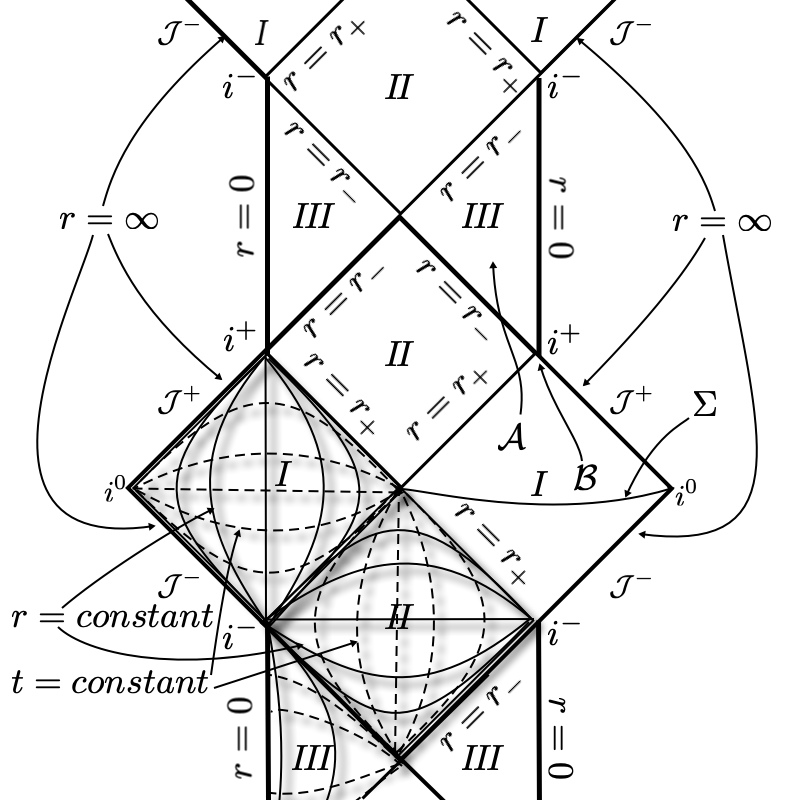}}
\caption{Penrose-Carter diagram for the maximally extended spacetime for regular center black hole with two horizons.
}\label{pencart}
\end{figure}

For the case that $\alpha > \frac{3\sqrt{3}}{2}$, our solutions depict a black hole with an outer event horizon ($r=r_+$) and inner Cauchy horizon ($r=r_-$). The Penrose-Carter diagram for this situation is depicted in Fig. \ref{pencart}. Future directed timelike paths can pass from asymptotically flat
regions $I$, across the horizon $r_+$,  through $I\! I$, $I\! I\! I$  and $I\! I$ and emerge in an asymptotically flat region $I$ in a different universe.  Furthermore, future directed causal curves from region $I$, crossing the horizon  $r_+$, may pass through regions  $I\! I$ and $I\! I\! I$ and cross the non-singular de Sitter space at $r=0$, wrapping around the universe until they exit into an asymptotically flat region $I$ via a region $I\! I$.  

In the Penrose-Carter diagram, the observer $\mathcal{B}$, remaining outside the horizon $r_+$ and traveling along the timelike worldline to future infinity at $i^+$, observes that particles $\mathcal{A}$ approaching the horizon $r_+$ slow and become infinitely redshifted.
An observer $\mathcal{A}$ crossing $r=r_-$ would observer the entire history of one of the regions $I$ in a finite time. Therefore, particles in the region $I$ would gain infinite blue shift as they approached future infinity $i^+$. Hence small perturbations of initial data on the spacelike surface $\Sigma$ would seemingly lead to catastrophic singularity creation on the surface $r=r_-$ (although actual infinities could perhaps be ameliorated by a UV cutoff at the model scale parameter $\ell$). 

 %%%%%%%%%%%%%%%%%

%=====================================================================
%=====================================================================
%=====================================================================
\subsection{Energy momentum tensor and energy conditions}\label{EnergyConditions}
%=====================================================================
%=====================================================================%=====================================================================

\begin{figure}[H]
\centerline{\includegraphics[width=4.75in]{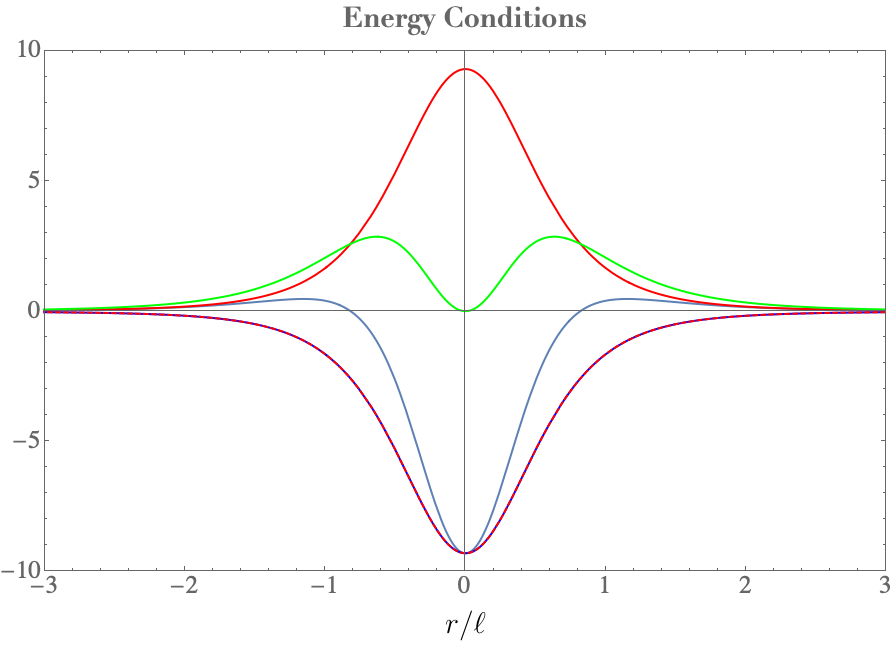}}
\caption{Plot of energy density $\rho$ (solid red), negative of the energy density $-\rho$ (dashed red), $\rho +p_\perp$ (green), $p_\perp$ (solid blue), and $p_\parallel$ (dashed blue). Model parameters are $m=1$, $\ell = .5$, $\Lambda = .05$. Since $p_\parallel=-\rho$, the dashed red and dashed blue curves are coincident.}
\label{energyconditionsplot}%Figure labels seem to need to go after the caption for the referencing to work IN ADDITION TO NOT INCLUDING A FILE EXTENSION
\end{figure}

Here, we consider the null, weak, dominant, and strong energy conditions for the T-dual-de Sitter black hole\footnote{Though we consider all of the energy conditions, the weak energy condition was examined in \cite{Mars:1996} for two asymptotically Schwarzschild regular black holes, and was expanded on in \cite{Borde:1996df} for a survey of similar black holes, while \cite{Ma:2015gpa} considered the weak energy condition for an asymptotically Reissner-Nordstrom regular black hole solution.}. For a typical solution having both black hole and cosmological event horizons, outside of the black hole we have
\be
2m <\frac{\left(\ell^2+r^2\right)^{3/2} \left(3 - \Lambda  r^2\right)}{3 r^2}\,,
\ee
and the energy density is $\rho = - T^t{}_t$. The pressure is given parallel to the radial coordinate by $p_\parallel = T^r{}_r$ and perpendicular to the radial coordinate $p_\perp = T^\theta{}_\theta= T^\phi{}_\phi$.
From the Einstein equation $G^\mu{}_\nu = 8 \pi G_N T^\mu{}_\nu$:
\bea
 \rho &=& \frac{6 \ell^2 m}{ 8 \pi G_N \left(\ell^2+r^2\right)^{5/2}}+\frac{\Lambda}{8 \pi G_N}, \\
p_\parallel &=& -\rho , \label{ppareq}\\
p_\perp &= &\frac{3 m  \ell^2 \left(3 r^2-2 \ell^2\right)}{ 8 \pi G_N \left(\ell^2+r^2\right)^{7/2}}-\frac{\Lambda}{8 \pi G_N}  \,.
\eea
Note that the null energy conditions (NEC) requires both $\rho +p_\parallel  \geq 0$  and $\rho +p_\perp  \geq 0$ for all values of the parameters. The first condition is clearly satisfied for all values of $\ell, \,\Lambda, \,m$ and $r$. In addition, 
\begin{equation}
\rho +p_\perp = \frac{15 \ell^2 m r^2}{8 \pi G_N \left(\ell^2+r^2\right)^{7/2}} \,,
\end{equation}
which is  non-negative for all parameter values we consider.  We have therefore shown that the NEC holds for all values of parameters. 
For the following discussion regarding energy conditions we refer the reader to Fig. \ref{energyconditionsplot}.

Inside any horizon we have 
\be
2m >\frac{\left(\ell^2+r^2\right)^{3/2} \left(3 - \Lambda  r^2\right)}{3 r^2}\,.
\ee
Beyond any horizon the $r$ coordinate becomes timelike and the $t$ coordinate spacelike. We therefore have that $\rho = - T^r{}_r$, $p_\parallel = T^t{}_t$ and $p_\perp = T^\theta{}_\theta= T^\phi{}_\phi$. Remarkably, for our spacetime $T^r{}_r= T^t{}_t$, and the expressions for $\rho$ and $p$ remain unchanged. We therefore deduce that the NEC is satisfied for all ranges of the $r$ coordinate. In addition, the weak energy condition (WEC) is satisfied since $\rho \ge 0$ and $\rho +p_\perp  \ge 0$ and $\rho +p_\parallel  \ge 0$. The dominant energy condition (DEC) can also be shown to hold: $\rho \ge 0$ , $p_\perp \in [-\rho,\rho]$ and $p_\parallel \in [-\rho,\rho]$. From (\ref{ppareq}), we see that the strong energy condition (SEC) is violated since, for certain values of parameters: $\rho + p_\parallel +p_\perp \ngeq 0$. We note that all of these findings are consistent with expectations from \cite{WiseSphericallySymCDC}, where energy conditions are discussed for nonsingular metrics in the context of the double copy.

%=====================================================================
%=====================================================================
\subsubsection{The nonsingular black hole from a magnetic monopole}\label{monopolegravsourceSection}
%=====================================================================
%=====================================================================
 
 In \cite{AyonBeato}, it was shown that the Bardeen black hole can be considered as an exact solution in general relativity with a source coming from nonlinear electrodynamics (NLED). A generic feature of NLED models is that gauge field quantities are finite everywhere, including the position of a point source. This may have motivated the authors of \cite{AyonBeato} to work towards understanding the Bardeen black hole as being sourced by a monopole from some NLED theory.
 
Specifically, the source is a magnetic monopole with charge $g$, with a Lagrangian given by
\begin{equation}\label{NLEDlagrangian}
\mathcal{L}=\frac{3m}{|g|^3}\Bigg(\frac{\sqrt{2g^2F}}{1+\sqrt{2g^2F}}\Bigg)^{5/2},
\end{equation}
where $F\equiv\frac{1}{4}F_{\mu\nu}F^{\mu\nu}$ and $m$ is the black hole mass. The ansatz for the gauge field and metric 
\begin{equation}\label{monopolemet}
-g_{tt}(r)=1-\frac{2M(r)}{r}, \ \ \ \ \ \ \ \ F_{\mu\nu}=2\delta^\theta_{[\mu}\delta^\phi_{\nu]}B(r,\theta),
\end{equation}
become solutions to the Einstein-NLED equations when
\begin{equation}
M(r)=\frac{mr^3}{(r^2+g^2)^{3/2}}, \ \ \ \ \ \ \ B(\theta)=g\sin\theta.
\end{equation}
This form matches the T-duality black hole, and produces all of the same curvature quantities in our Appendix A with the substitution $\ell\rightarrow g$ (and with $\Lambda=0$). 

Asymptotically, the metric (\ref{monopolemet}) behaves as
\begin{equation}
-g_{tt}(r)=1-\frac{2m}{r}+\frac{3mg^2}{r^3}+O(r^{\text{-}5}).
\end{equation}
This is clearly different to the Reissner-Nordstrom solution, which assumes a static electric monopole source of the form $F_{\mu\nu}\propto\frac{q}{r^2}\delta^{t}_{[\mu}\delta^r_{\nu]}$ in Maxwell theory with $\mathcal{L}(F)=-F$, and results in the metric $-g_{tt}(r)=1-\frac{2m}{r}+\frac{q^2}{r^2}$ (up to the dimensionful factors). Comparatively, we see that the nonlinear magnetic monopole's contribution to the metric falls off more rapidly than its linear, electric monopole counterpart. We will return to the discussion of the sources in the context of the double copy.

%=====================================================================
%=====================================================================
%===================================================================== 
   \subsection{Orbital equations}\label{Orbits}
%=====================================================================
%=====================================================================
%=====================================================================

We move on to consider the basic details of circular orbits in the T-dual-de Sitter black hole background. A full treatment of all the orbital properties of the Bardeen spacetime (without a cosmological constant) can be found in \cite{Zhou:2011aa}.

 Metric compatibility and the geodesic equation imply that the following is constant along particle trajectories in our background spacetime:
 \be\label{epseq}
 \epsilon =- g_{\mu\nu} \frac{dx^\mu}{d\lambda} \frac{dx^\nu}{d\lambda}  \,,
 \ee
 where $\epsilon = 1$ for timelike, massive particles and $\epsilon =0$ for null, massless particles. Without loss of generality we consider motion in the equatorial plane with
 $\theta = \pi/2$. Expanding (\ref{epseq})  gives
 \be
 g_{tt} \left( \frac{dt}{d\lambda}  \right)^2 -  g_{rr} \left( \frac{dr}{d\lambda}  \right)^2 -  r^2 \left( \frac{d\phi}{d\lambda}  \right)^2 = \epsilon \,.
 \ee
 We consider the timelike Killing vector (\ref{tlkv}), and  spacelike Killing vector for the angular coordinate:
 \be
  R^\mu = (\partial_\phi)^\mu = (0,\, 0,\, 0,\, 1) \,,
 \ee
 which lead to the conserved energy $E$ and angular momentum $L$ for photons (conserved energy and angular momentum per unit mass for massive particles):
 \be
 E = - K_\mu \frac{dx^\mu}{d\lambda}  =  \left(1 -\frac{2 m r^2}{\left(r^2 +\ell^2 \right)^{3/2}}-\frac{\Lambda  r^2}{3} \right)  \frac{dt}{d\lambda}  \,,
 \ee
 \be
 L = R_\mu \frac{dx^\mu}{d\lambda}  = r^2  \frac{d\phi}{d\lambda}  \,.
 \ee
 In terms of the conserved quantities we have:
 \be 
 \left( \frac{dr}{d\lambda}  \right)^2 =  E^2 - \left(1 -\frac{2 m r^2}{\left(r^2 +\ell^2 \right)^{3/2}}-\frac{\Lambda  r^2}{3} \right) \left(\epsilon + \frac{L^2}{r^2} \right ) \,,
 \ee
 yielding effective orbital potentials:
 \bea\label{efpot}
  V_\epsilon(r) &=&  - \frac{1}{2} g_{tt}  \left( \frac{L^2}{r^2}+\epsilon \right) \, \\
% V = \frac{1}{2}  \left(1 -\frac{2 m r^2}{\left(r^2 +\ell^2 \right)^{3/2}}-\frac{\Lambda  r^2}{3} \right) \left(\epsilon + \frac{L^2}{r^2} \right )
&& = \frac{\epsilon}{2} -\frac{m r^2 \epsilon }{\left(\ell^2+r^2\right)^{3/2}}  + \frac{L^2}{2 r^2} -\frac{L^2 m}{\left(\ell^2+r^2\right)^{3/2}} -\frac{\Lambda  L^2}{6}-\frac{ \Lambda \epsilon} {6} r^2  \,.
 \eea
 Ignoring the de Sitter complications of the later terms the general relativistic corrections to the Newtonian orbital analysis result in the fourth term in (\ref{efpot}) with $\ell =0$, i.e.  a term which dominates $V$ at small $r$ that is proportional to $- 1/r^3$.
 Perhaps the most significant difference between the above potential and that derived from the familiar Schwarzschild solution occurs at small values of the radial coordinate. Due to the absence of the singularity in the black hole solution (\ref{gfunc}), the fourth term in $V$ no longer diverges at $r=0$; thus,  the Newtonian angular momentum barrier term 
 (third term) in $V$ dominates the potential at small $r$ which is proportional to $+1/r^2$. Due to the complexity of the metric under consideration we apply numerical analysis to the remainder of the orbital analysis.

 %=====================================================================
 %=====================================================================
 %\subsection{Circular orbits}
 %=====================================================================
 %=====================================================================

%=====================================================================
 \subsubsection{Photon orbits}
 %=====================================================================
 
 The orbital dynamics of massless particles or photons traveling along null paths are governed via the effective potential with $\epsilon =0$:
 \be\label{pot0}
 V_0(r) =   \frac{L^2}{2 r^2} -\frac{L^2 m}{\left(\ell^2+r^2\right)^{3/2}} -\frac{\Lambda  L^2}{6}  \,.
 \ee
 Circular photon orbits are located at $r=r_c$, where $r_c$ is obtained via $V^\prime_0(r_c)=0$, where
 \be\label{photonorbits}
V^\prime_0(r_c)= \frac{3 L^2 m r_c}{\left(\ell^2+r_c^2\right)^{5/2}}-\frac{L^2}{r_c^3}\,.
 \ee
 The circular photon orbits  occur at maxima (unstable) and minima (stable) of the potential (\ref{pot0}) (see Fig.~7).  For simplicity we focus on the case where $\Lambda =0$ (see e.g.~\cite{Zhou:2011aa}). We find outside of the black hole outer horizon $r_+$, unstable circular orbits for photons exist at the maxima of the potential. The effective potential vanishes at the outer and inner horizons $r_\pm$. Interestingly, and unlike the ordinary (singular) Schwarzschild solution, we find stable circular orbits for photons exist in between the outer and inner horizons, $r_+$ and $r_-$.
 %\newpage
 \begin{figure}\label{pot0.png}
\centerline{\includegraphics[width=4.75in]{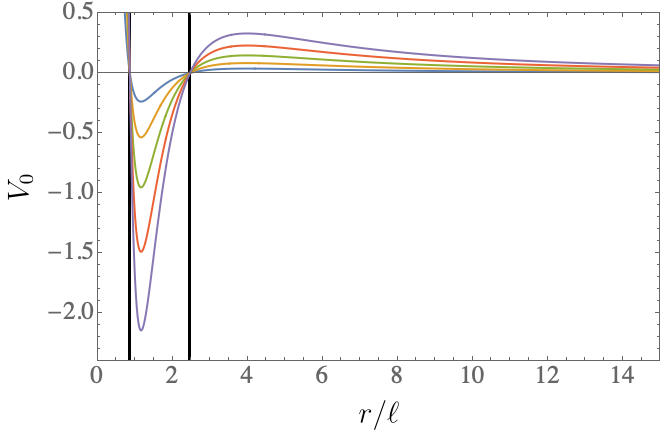}}
\caption{Plot of null $\epsilon=0$ effective potential (\ref{pot0}) for various values of angular momentum ($L=6$ purple, $L=5$ red, $L=4$ green, $L=3$ orange, $L=2$ blue.) The solid black lines show the location of the inner and outer horizons. 
}
\end{figure}

%========================================================================
%========================================================================
 %=====================================================================
  \subsubsection{Massive particle orbits}
 %=====================================================================
  %========================================================================
  %========================================================================
  
 The orbital dynamics of massive particles traveling along timelike paths are governed via the effective potential with $\epsilon =1$:
 \be\label{pot1}
 V_1(r) =  \frac{1}{2} \left(\frac{L^2}{r^2}+1\right) \left(1-\frac{2 m r^2}{\left(\ell^2+r^2\right)^{3/2}}-\frac{\Lambda  r^2}{3}\right)  \,.
 \ee
 Circular massive particle orbits are located at $r=r_c$, where $r_c$ is obtained via $V^\prime_1(r_c)=0$, where
 \be
V^\prime_1(r_c)= L^2 \left(\frac{3 m r_c}{\left(\ell^2+r_c^2\right)^{5/2}}-\frac{1}{r_c^3}\right)+\frac{m r_c \left(r_c^2-2 \ell^2 \right)}{\left(\ell^2+r_c^2\right)^{5/2}}-\frac{\Lambda  r_c}{3}\,.
 \ee
 The circular massive particle orbits  occur at maxima (unstable) and minima (stable) of the potential (\ref{pot1}) (see Fig.~8).  For simplicity we focus on the case where $\Lambda =0$. We find for sufficiently high $L$ it is possible to have outside of the black hole outer horizon $r_+$, one unstable circular orbit radius  exists at the maxima of the potential and one stable circular radius at large $r$. The effective potential vanishes at the outer and inner horizons $r_\pm$. Again, interestingly, and unlike the ordinary (singular) Schwarzschild solution, we find stable circular orbits for massive particles exist in between the outer and inner horizons, $r_+$ and $r_-$.
 
  %\newpage

 \begin{figure}[H]\label{pot1.png}
\centerline{\includegraphics[width=4.75in]{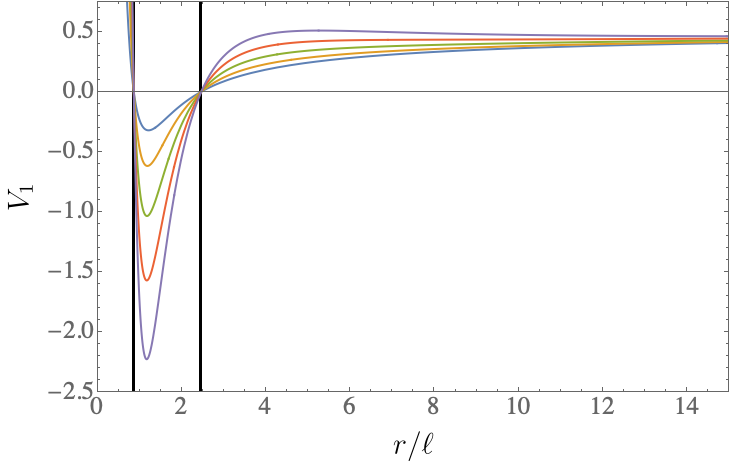}}
\caption{Plot of timelike massive particles with $\epsilon=1$ effective potential (\ref{pot1}) for various values of angular momentum ($L=6$ purple, $L=5$ red, $L=4$ green, $L=3$ orange, $L=2$ blue.) The solid black lines show the location of the inner and outer horizons. 
}
\end{figure}

 %=====================================================================
 %=====================================================================
 %===================================================================== 
 \section{Non-singular double copy}\label{NSBHdoublecopy}
 %=====================================================================
 %=====================================================================
 %=====================================================================

The double copy story originates with \cite{Bern1}, where it was observed that complicated graviton scattering amplitudes can be obtained from simpler gauge field scattering amplitudes. Briefly summarizing, one starts with a Yang-Mills amplitude $\mathcal{A}^{\text{YM}}$ and exchanges a color factor $c_i$ with a kinematic numerator $\tilde{n}_i$ to obtain the graviton amplitude $\mathcal{M}^{\text{grav}}$. Schematically,

\begin{equation}\label{amplitudeDC}
\mathcal{A}^{\text{YM}}\sim\sum_k \frac{n_k c_k}{\text{propagators}} \ \ \ \ \ \ \ \overset{c_k\rightarrow \tilde{n}_k}{\xrightarrow{\hspace*{2cm}}}  \ \ \ \ \ \ \sum_k \frac{n_k \tilde{n}_k}{\text{propagators}}\sim \mathcal{M}^{\text{grav}},
\end{equation}
%====================
where the sum is over all three-point vertex graphs, the $n_k$ are the kinematic numerators associated with each graph, and the $c_k$ are the color factors that satisfy a Jacobi identity of the form $c_i+c_j+c_k=0$. The second set of kinematic numerators $\tilde{n}_k$ are also organized to also satisfy an identical Jacobi identity.

There is also a `zeroth copy' in the amplitudes story, where, starting with the left hand side of (\ref{amplitudeDC}), replacing the kinematic numerators $n_i$ with a second set of color factors $\tilde{c}_i$ builds scalar amplitudes of the form
\begin{equation}\label{scalaramp} 
\mathcal{A}^{\text{scalar}}\sim\sum_k\frac{c_k\tilde{c}_{k}}{\text{propagators}},
\end{equation}
for bi-adjoint scalars $\phi^{aa'}$.

As a basic example, pure (non-supersymmetric) Yang-Mills mills theory double copies to general relativity coupled to a two-form field and a dilaton. These relations are perturbative statements, however the authors of \cite{MonteiroMainCDC} showed that a double copy procedure can be applied to relate exact solutions in general relativity to exact solutions in the $U(1)$ sector of Yang Mills, aptly named the classical double copy.

The first realization\footnote{A second version of the classical double copy appeared in \cite{LunaTypeD}, called the Weyl double copy, where the relationship between the gravity side and gauge side is built using the symmetric four-index Weyl spinor and the spinor forms of the gauge field strengths. More recently, \cite{Elor:2020nqe} presented a map for self-dual solutions using the Newman-Penrose formalism. Both of these works have shown equivalences to the Kerr-Schild double copy.} of the classical double copy relies on a spacetime admitting a Kerr-Schild metric, 
\begin{equation}\label{KerrSchild}
g_{\mu\nu}=\eta_{\mu\nu}+\phi k_\mu k_\nu,
\end{equation}
where $k_\mu$ is null with respect to both the flat background and the full metric, $g^{\mu\nu}k_\mu k_\nu=\eta^{\mu\nu}k_\mu k_\nu=0, $ and $\phi=\phi(x^\mu)$ is a scalar function. The statement relating the gravity solution to a gauge theory solution is \textit{if} $g_{\mu\nu}$ \textit{satisfies the Einstein equations, then the gauge field}
\begin{equation}\label{singlecopy}
A^a_\mu=\phi c^a k_\mu
\end{equation}
\textit{satisfies the} $U(1)$ \textit{sector of Yang-Mills over the background} $\eta_{\mu\nu}$. The gauge field (\ref{singlecopy}) is referred to as the single copy. The $c^a$ is a color vector carrying the Non-Abelian index, and analogously to $c_k\rightarrow \tilde{n}_k$ in (\ref{amplitudeDC}), in the classical double copy, the graviton is obtained by exchanging the color vector with a copy of the null vector:
\begin{equation}
 A^a_\nu=\phi c^a k_\nu \ \ \ \overset{c^a\rightarrow k_\mu}{\xrightarrow{\hspace*{2cm}}} \ \ \ \phi k_\mu k_\nu = h_{\mu\nu}.
\end{equation}
Thus, the gauge field can be thought of as being a `square root' of the graviton. Since the $U(1)$ sector of Yang-Mills is Abelian, the $c^a$ play a passive role in the single copy equations and we generally won't write it with the gauge fields. Further, the scalar function $\phi$ plays the role of the zeroth copy in the classical story, and satisfies a (possibly sourced) wave equation over $\eta_{\mu\nu}$.

In the Kerr-Schild coordinates (\ref{KerrSchild}), the mixed Ricci tensor reads
\begin{equation}
R^\mu_{ \ \nu}=\frac{1}{2}\Big(\partial^\mu \partial_\alpha(\phi k^\alpha k_\nu)+\partial_\nu\partial^\alpha(\phi k_\alpha k^\mu)-\partial^2(\phi k^\mu k_\nu)\Big).
\end{equation}
If we specialize to the case where the metric is completely time independent, $\partial_0\phi=\partial_0 k_\mu=0$, then we may write the $R^\mu_{ \ 0}$ component of the mixed Ricci tensor associated with (\ref{KerrSchild}) as
\begin{equation}
R^\mu_{ \ 0}=-\frac{1}{2}\partial_\nu\big(\partial^\mu(\phi k^\nu)-\partial^\nu(\phi k^\mu)\big)=-\frac{1}{2}\partial_\nu F^{\mu\nu},
\end{equation}
which illustrates the connection between curvature on the gravity side and the Maxwell equations via the ansatz (\ref{singlecopy}). Here, the derivatives $\partial_\mu$ are taken over the flat background $\eta_{\mu\nu}$. Moreover, in the time independent case, the zeroth copy is related to the mixed Ricci tensor by
\begin{equation}\label{RicciAndZeroth}
R^0_{ \ 0}=\frac{1}{2}\partial^2\phi.
\end{equation}

In the presence of a source where $\partial_\nu F^{\mu\nu}=4\pi J^\mu,$ we have that $R^\mu_{ \ 0}=-2\pi J^\mu.$ Instead in terms of the Einstein tensor, this is
\begin{equation}\label{sourcerelation}
G^\mu_{ \ 0}+\frac{1}{2}R\delta^\mu_{ \ 0}=-2\pi J^\mu \ \ \ \ \ \Rightarrow \ \ \ \ \ T^\mu_{ \ 0}+\frac{1}{2}R\delta^\mu_{ \ 0}=-2\pi J^\mu,
\end{equation}
where we use the Einstein equations\footnote{We follow the convention $m_p^2=\frac{1}{8\pi G}=1$.} to write the second relation. (\ref{sourcerelation}) shows that the gauge field source can be thought of as coming from both the gravitational source and the spacetime curvature. We will revisit the relation in section \ref{singlecopysourcediscussion}. We next write arbitrary spherically symmetric metrics in the form (\ref{KerrSchild})  and analyze the non-singular black hole solutions.

%========================================================================
%=====================================================================
%===================================================================== 
\subsection{Kerr-Schild single copy}\label{KSscSection}
%=====================================================================
%=====================================================================
%======================================================================== 
To write the metric (\ref{metricg}) in the Kerr-Schild form (\ref{KerrSchild}), we make the coordinate transformation
\begin{equation}\label{difftoKerrSchild}
    d\tau=dt-(1+g_{tt}^{-1})dr .
\end{equation}
This indeed has the correct form over the Minkowski background, $g_{\mu\nu}=\eta_{\mu\nu}+\phi k_\mu k_\nu$, with $\eta_{\mu\nu}=\text{diag}(-1,1,r^2,r^2\sin^2\theta), $ $k_\mu=(1,1,0,0) $ and  $\phi(r)=1+g_{tt}(r).$ The single copy gauge field is $A_{\mu}=\phi(r)(1,1,0,0),$ which subject to the gauge transformation
\begin{equation}
A_{\mu}\rightarrow A_{\mu}+\partial_\mu\chi, \ \ \ \ \chi=-\int\phi(r')dr' \ \ \ \ \Rightarrow \ \ \ \ A_{\mu}=\phi(r)(1,0,0,0),
\end{equation}
leaves us with a Coulomb-type solution. For the T-dual-de Sitter black hole, the scalar profile is
\begin{equation}\label{scalar}
    \phi(r)=\frac{2mr^2}{(r^2+\ell^2)^{3/2}}+\frac{\Lambda}{3}  r^2.
\end{equation}

We examine the electric fields associated with the gauge theory, defined covariantly as
\begin{equation}
E_\mu=F_{\mu\nu}K^\nu,
\end{equation}
where $K^\mu$ is the timelike Killing vector (\ref{tlkv}). Computing the field strength $F_{\mu\nu}=\partial_{[\mu}A_{\nu]}$ over the flat Minkowski background, we get one nonzero component, $F_{\tau r}$, which, contracted with $K^\mu$ gives a radial electric field
\begin{equation}\label{electricfield}
    E_r(r) = \frac{2mr(2\ell^2-r^2)}{(r^2+\ell^2)^{5/2}}+\frac{2\Lambda }{3}r \,.
\end{equation}
The associated charge density $J^0=\rho_c$ obtained from $\partial_\nu F^{\mu\nu}=4\pi J^\mu$ is
\begin{equation}\label{rhocharge}
4\pi \rho_c(r) = \frac{6 m \ell^2  \left(2\ell^2-3  r^2\right)}{\left(r^2+\ell^2\right)^{7/2}}+2 \Lambda \,.
\end{equation}

From (\ref{scalar}), (\ref{electricfield}), and (\ref{rhocharge}), it is easy to see that every relevant quantity in the gauge theory is completely non-singular for all values of $r$ in the range $r\in(-\infty,\infty)$, including the origin:
\begin{equation}
\lim_{r\rightarrow 0}A_\mu = 0, \ \ \ \ \ \ \ \ \ \lim_{r\rightarrow 0}F_{\mu\nu}=0, \ \ \ \ \ \ \ \ \ \lim_{r\rightarrow 0}\partial_\nu F^{\mu\nu}=\text{constant}.
\end{equation}

Note that expanding the charge density around $r=0$ gives 
\begin{equation}\label{rhocAroundOrigin}
4\pi \rho_c(r)=\frac{12m}{\ell^3}+2\Lambda +O(r^2). 
\end{equation}
As was discussed in section \ref{sec:nsbh}, the metric limits to a de Sitter core as $r\rightarrow 0$, with length scale $\frac{1}{l^2}=\frac{\Lambda}{3}+\frac{2m}{\ell^3}$. Hence from (\ref{rhocAroundOrigin}), we find $\rho_c =  \frac{6}{l^2}+O(r^2)$, which is consistent with the expectation of a pure de Sitter single copy produced by a constant charge density \cite{LunaCurvedBckgrnd}.

Finally, observe that in the $r<0$ region of the spacetime, a positive (negative) charge density sources a negative (positive) electric field. This sign is the opposite of what we have defined as the normal convention in our universe. This feature can be traced to the first component of the Maxwell equations, $\partial_\nu F^{\mu\nu}\delta^\tau_\mu=4\pi J^\mu\delta^\tau_\mu$, which is just Gauss's law. In spherical coordinates, the non-covariant divergence (for $\vec{E}=E_r\hat{r}$) is $\vec{\nabla}\cdot\vec{E}=(\frac{2}{r}+\partial_r) E_r\propto\rho_c$. The operator $(\frac{2}{r}+\partial_r)$ is antisymmetric in $r\rightarrow -r,$ therefore if we imagine $\rho_c>0,$ it must be that $E_r<0$ when $r<0.$ 

\subsubsection{Single copy electric fields and charge density}\label{EsandRhos}

We now proceed to discuss the electric fields and charge densities associated with the $M,A,B,$ and $C$ classes of the T-dual-de Sitter metric. We  plot the dimensionless forms of each field quantity as $\tilde{E}_r=\ell E_r$ and $\tilde{\rho}_c=\ell^2\rho_c,$ where
\begin{equation}\label{dimlessEandrho}
\tilde{E}_r=\frac{\alpha\tr(2-\tr^2)}{(1+\tr^2)^{5/2}}+\frac{2\tL}{3}\tr, \ \ \ \ \ 4\pi \tilde{\rho}_c=\frac{3\alpha(2-3\tr^2)}{(1+\tr^2)^{7/2}}+2\tL,
\end{equation}
and $\tr=r/\ell$ as before.

%========================================================================
%========================================================================
%========================================================================
\paragraph{Classes \textbf{\textit{M}} and \textbf{\textit{A}}}
%========================================================================
%========================================================================
%========================================================================

The gauge theory quantities $E_r$ and $\rho_c$ are very similar for both the asymptotically Minkowski class of the T-Dual black hole and the T-Dual-de Sitter spacetime. By inspection to (\ref{dimlessEandrho}), the only difference between $\tL=0$ and $\tL\neq 0$ is shifting $\rho_c$ by a constant, while the electric field's contribution from $\tL$ amounts to a term linear in $\tr$ that dominates away from the origin. 

\begin{figure}[H]
\centering
\begin{subfigure}{.5\textwidth}
  \centering
  \includegraphics[scale=.35]{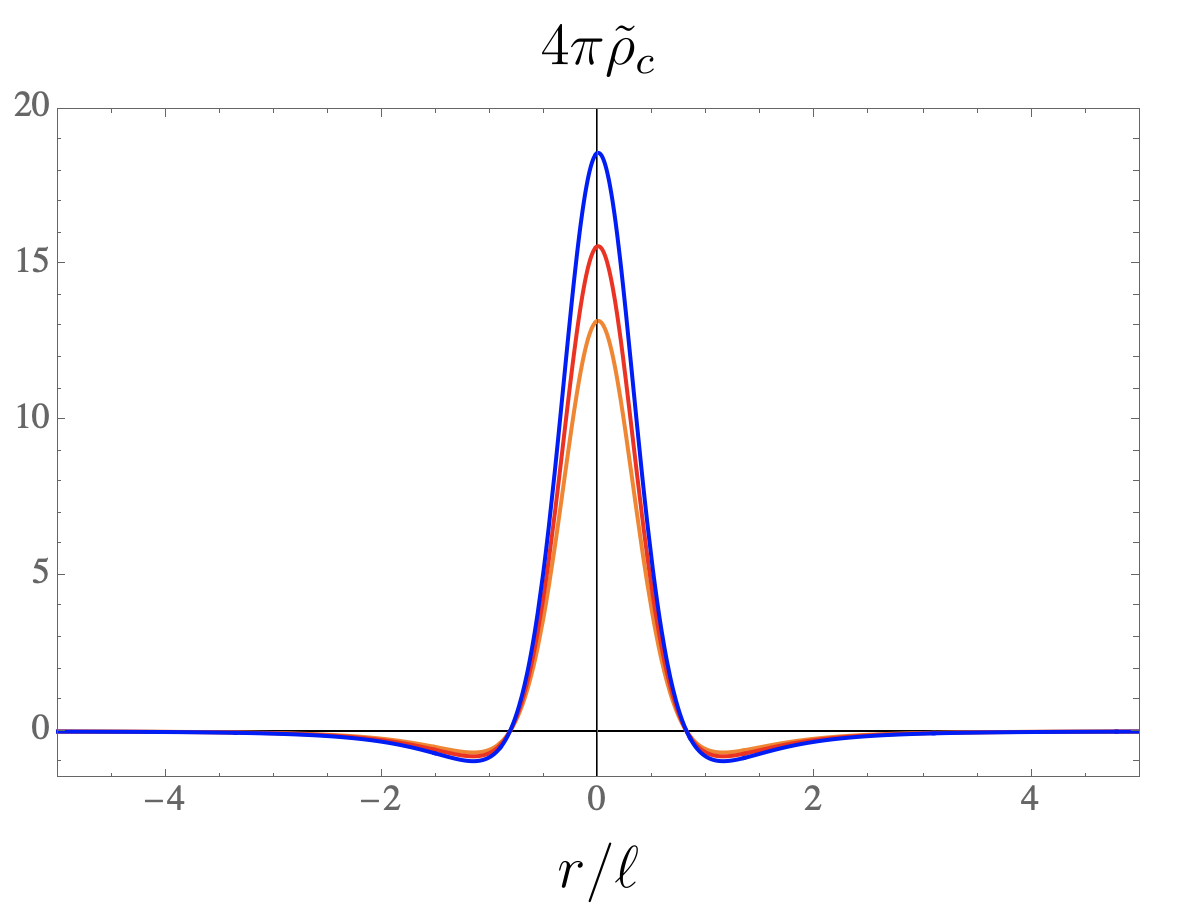}
\end{subfigure}%
\begin{subfigure}{.5\textwidth}
  \centering
  \includegraphics[scale=.35]{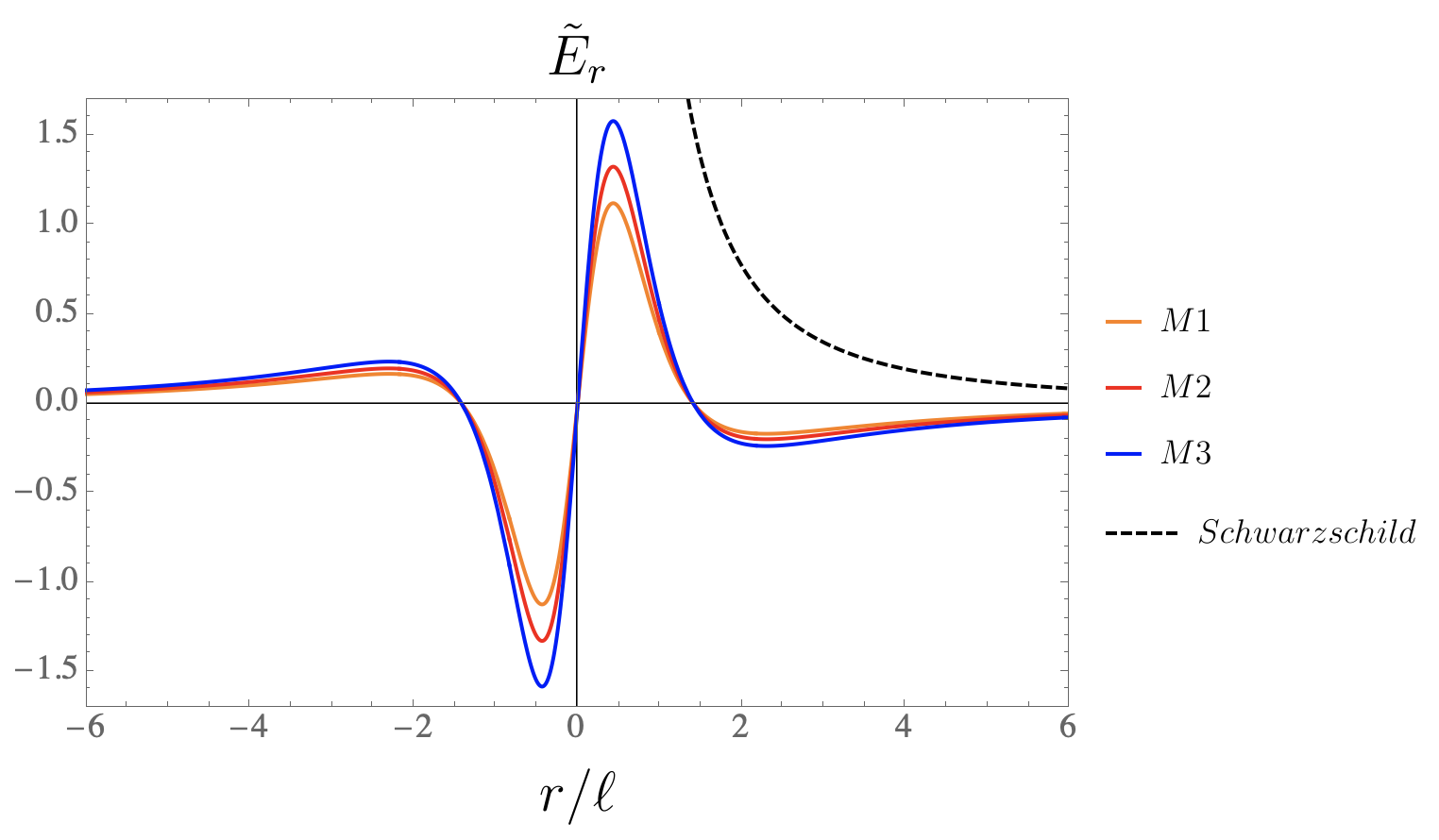}
\end{subfigure}
\caption{The dimensionless charge density and radial electric fields associated with the asymptotically Minkowski T-dual black hole solutions. The horizonless wormhole ($M1$) is shown in orange, the extremal black hole ($M2$) with two degenerate horizons in red, and the four horizon non-singular black hole in blue ($M3$). Each color/subclass on both of the plots above corresponds to the same parameter value as the $-g_{tt}$ curves in Fig. \ref{classMplot}. The electric field associated with the Schwarzschild solution, $E_r\sim m/r^2$, is shown in dashed black. The charge density for such a solution is given by a Dirac delta function and is not shown in the left panel.}
\label{rhoandEfigsForMclass}
\end{figure}

\begin{figure}[H]
\centering
\begin{subfigure}{.5\textwidth}
  \centering
  \includegraphics[scale=.35]{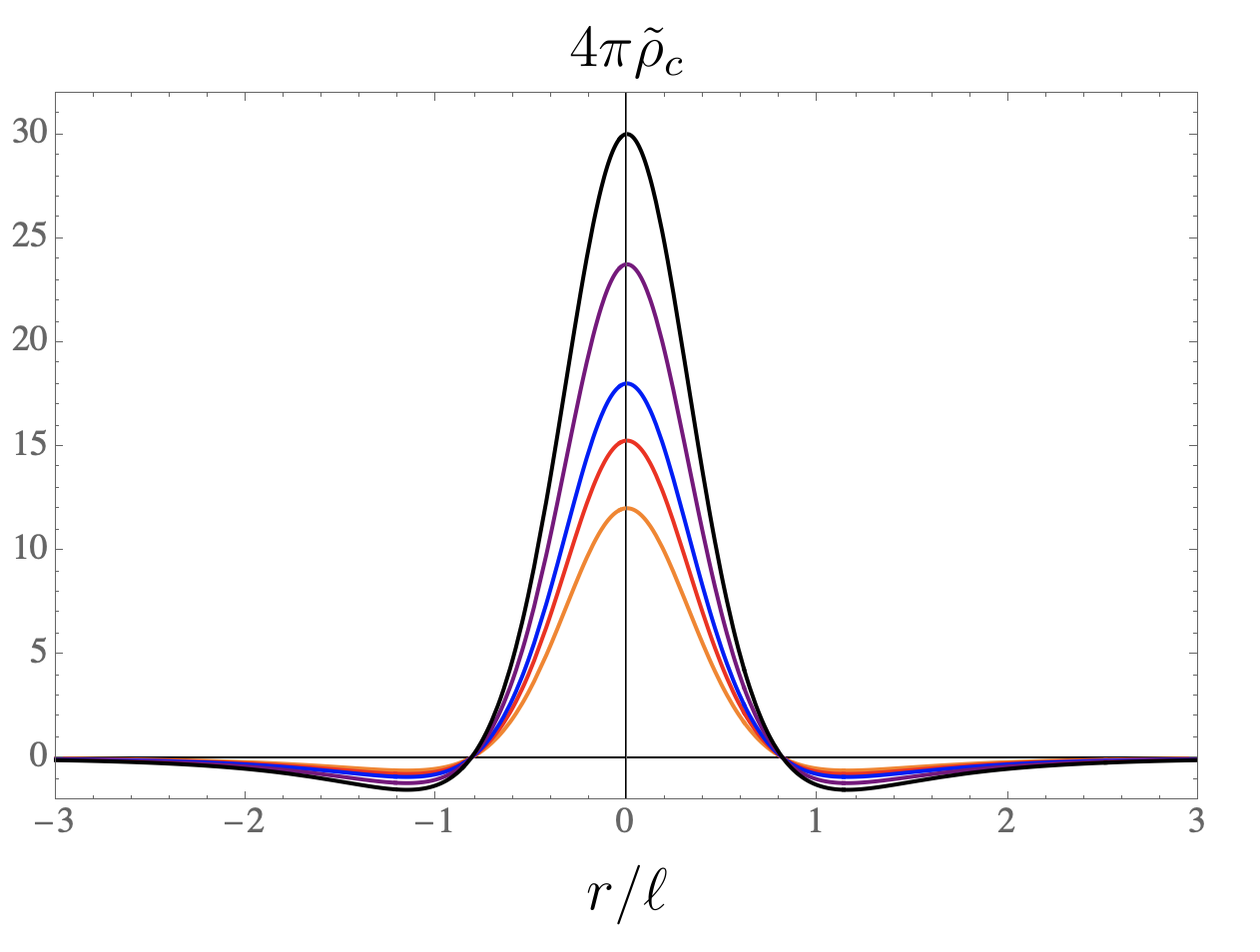}
\end{subfigure}%
\begin{subfigure}{.5\textwidth}
  \centering
  \includegraphics[scale=.35]{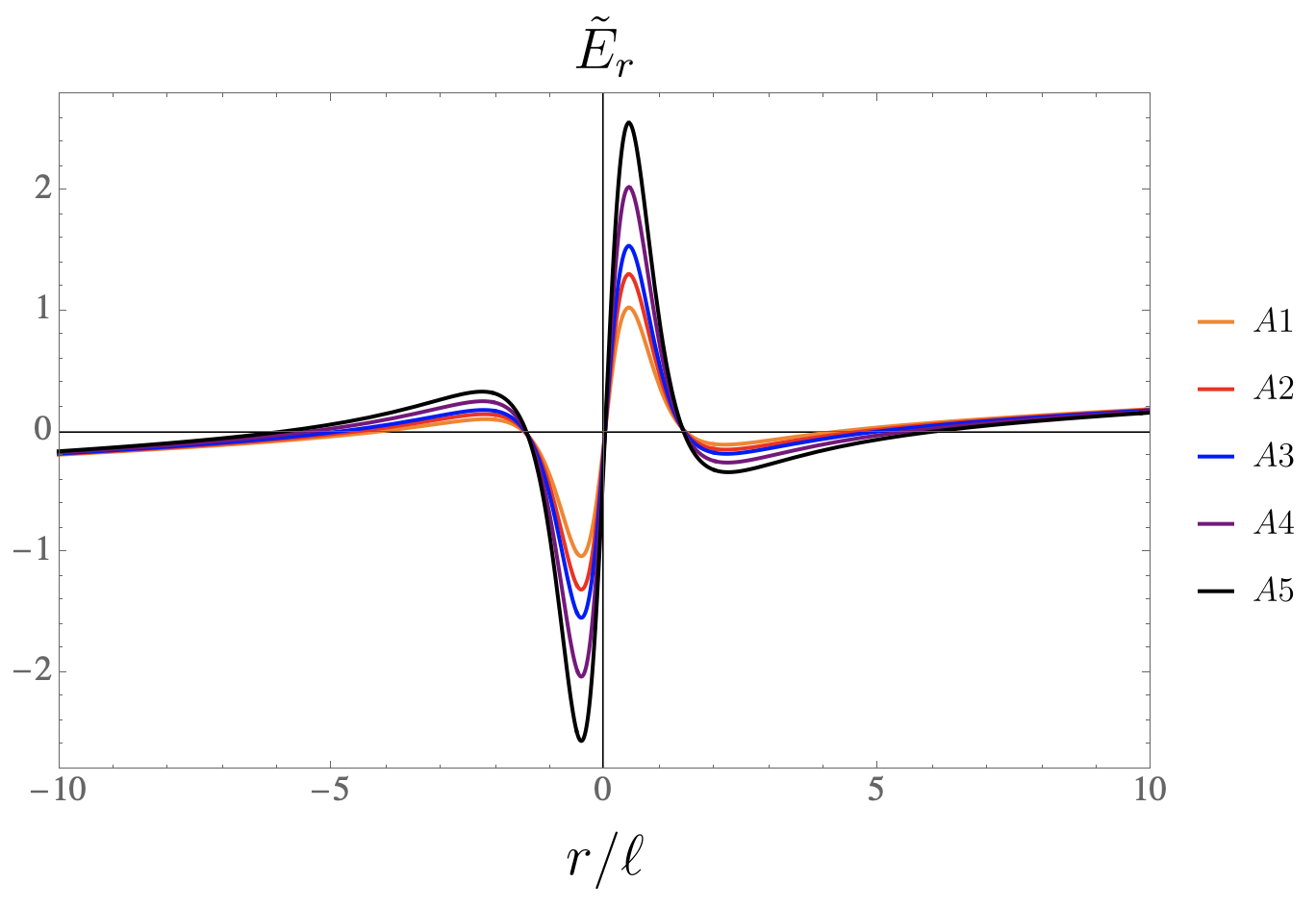}
\end{subfigure}
\caption{The charge density and radial electric fields associated with the T-Dual-de Sitter solution. The parameter values used the make each of the above curves are the same as those used to plot Fig. \ref{classAplot}. We have the wormhole with two cosmological horizons ($A1$) in orange, the degenerate black hole with two cosmological horizons ($A2$) in red, the six horizon spacetime ($A3$) in blue, the Nariai spacetime ($A4$) in purple, and the solution with two linear horizons ($A5$) in black. }
\label{rhoandEfigsForAclass}
\end{figure}

Interestingly, the electric field changes signs when approaching the origin from large $\pm\tr$. Due to the spherical symmetry of the spacetime, this sign change occurs on concentric 2-spheres centered at $\tr=0.$  This behavior is clearly visible in Figures \ref{rhoandEfigsForMclass} and \ref{rhoandEfigsForAclass}, as well as in the electric field equation (\ref{dimlessEandrho}). For class $A$, with $\tL\neq 0$ , the single copy electric field switches sign twice when approaching the origin, while for class $M$ where $\tL=0$ there is only one sign switch, as seen in Figures \ref{rhoandEfigsForMclass} and \ref{rhoandEfigsForAclass}. We will return to this point in section \ref{VanEfieldsAndHorizons}.

%========================================================================
%========================================================================
%========================================================================
\paragraph{Classes \textbf{\textit{B}} and \textbf{\textit{C}}}
%========================================================================
%========================================================================
%========================================================================

\begin{figure}[H]
\centering
\begin{subfigure}{.5\textwidth}
  \centering
  \includegraphics[scale=.35]{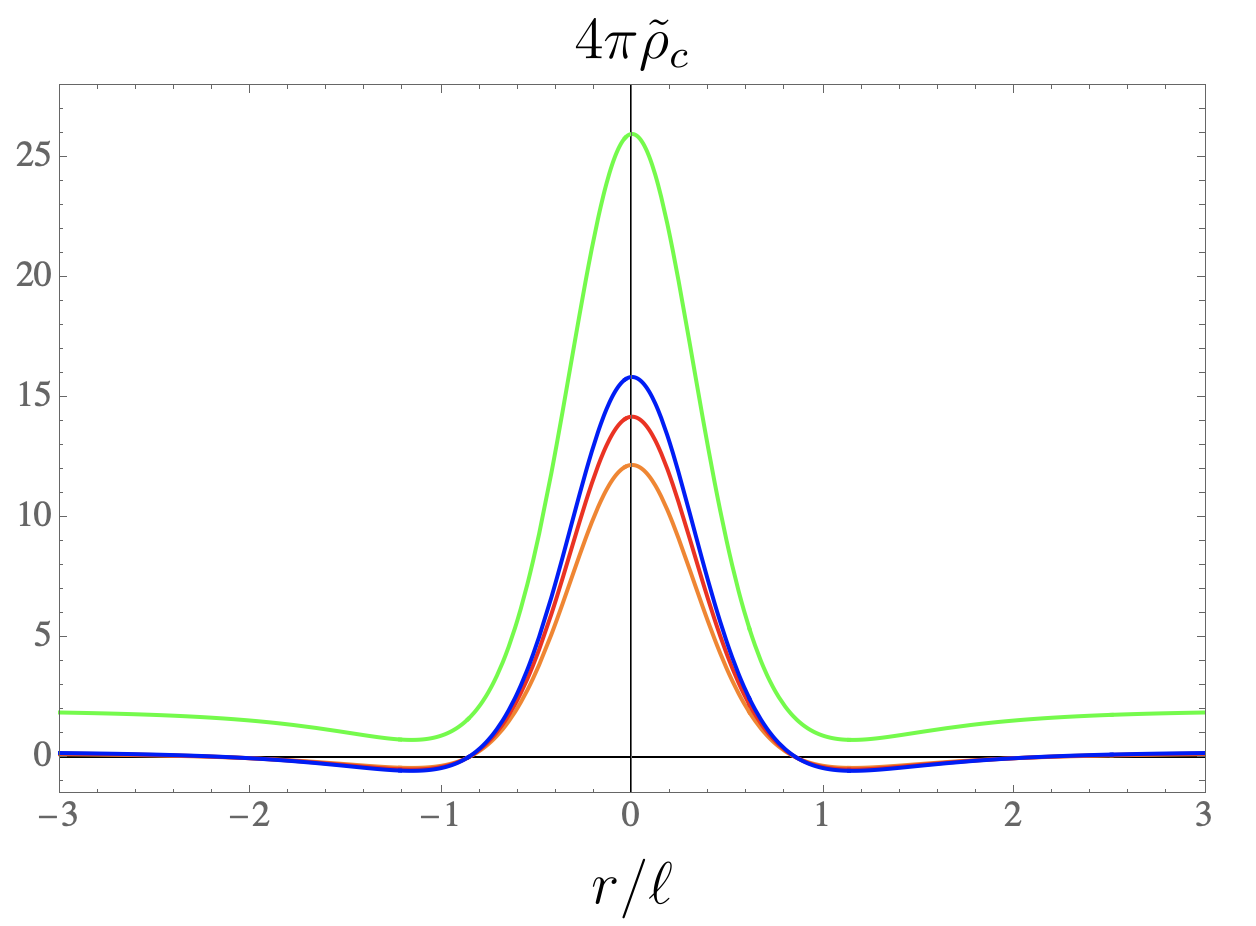}
\end{subfigure}%
\begin{subfigure}{.5\textwidth}
  \centering
  \includegraphics[scale=.35]{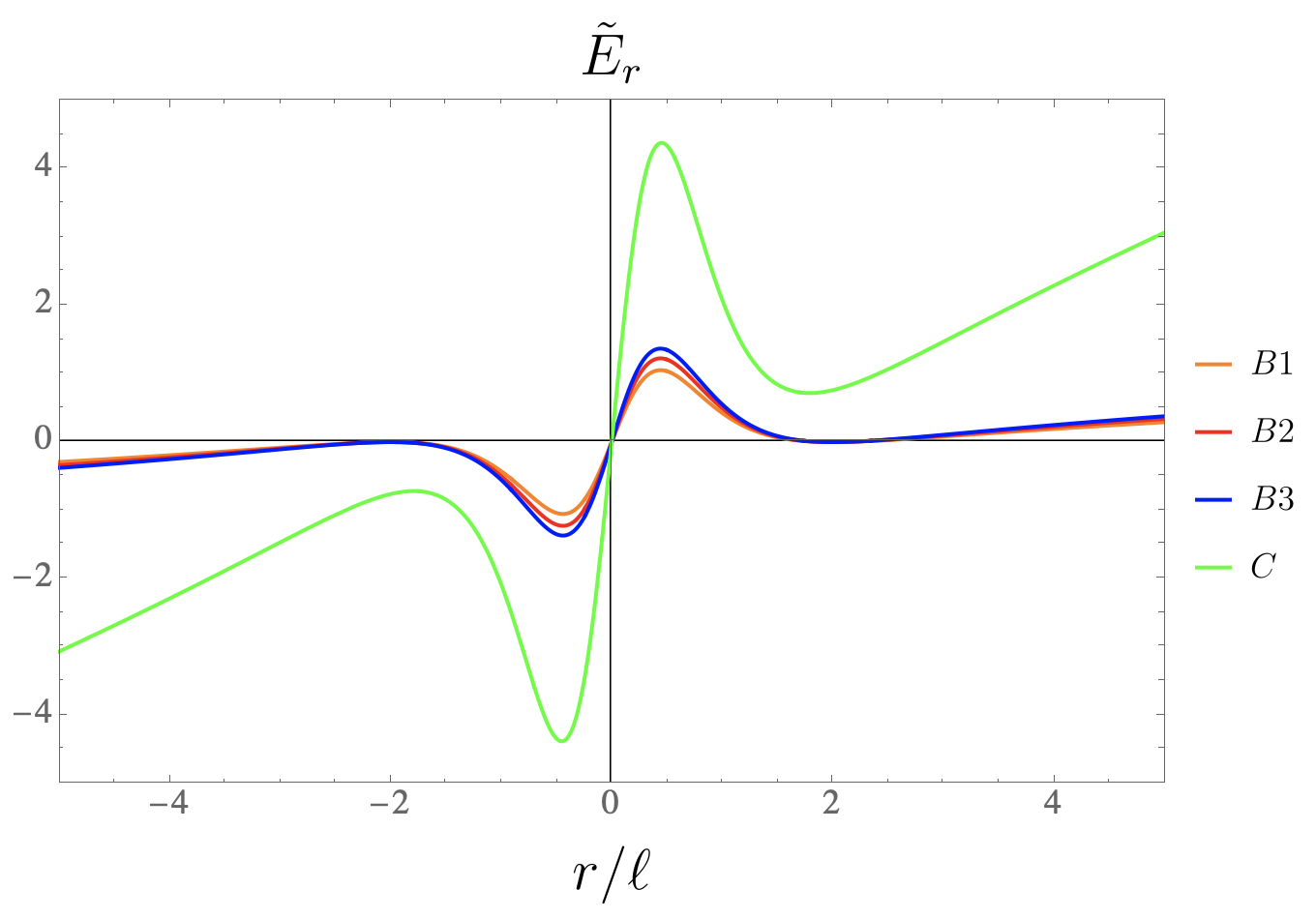}
\end{subfigure}
\caption{The charge density and radial electric fields associated with the T-Dual-de Sitter solution. The parameter values used the make each of the above curves are the same as those used to plot Fig. \ref{classesBCplot}. The triply degenerate horizon spacetime is shown in orange ($B2$), while the double linear horizon black holes are shown in orange, blue, and green ($B1$, $B3$, and $C$). We remind the reader that class $C$ solutions are such that $-g_{tt}$ has no local extrema, while the class $B$ solutions exhibit a single inflection point in each ($\tr>0$ and $\tr<0$) region.}
\label{rhoandEfigsForBCclass}
\end{figure}

As we alluded to in section \ref{AssDeSitGravSection}, the $B$ and $C$ class spacetimes are associated to gauge theories that behave slightly differently from the $M$ and $A$ classes. Mainly, instead of exhibiting sign changes when approaching the origin from positive or negative $\tr$, the electric field either vanishes \textit{without} changing signs (class $B$) or does not change signs at all (class $C$). These features are apparent in Fig. \ref{rhoandEfigsForBCclass}.

%========================================================================
%========================================================================
%========================================================================

\subsubsection{Discussion of the sources}\label{singlecopysourcediscussion}

%========================================================================
%========================================================================
%========================================================================

We briefly described in section \ref{monopolegravsourceSection} that the Bardeen black hole is an exact solution to the Einstein equations coupled to a nonlinear electrodynamic theory. Let us consider this in the context of our single copy gauge field source. In (\ref{sourcerelation}), we have only the $tt$-component being nonzero, so
\begin{equation}
T^\mu_{ \ 0}+\frac{1}{2}R\delta^\mu_{ \ 0}=-2\pi J^\mu\ \ \ \ \ \ \ \Rightarrow \ \ \ \ \ \ \ \ -\rho_{\text{NLED}}+\frac{1}{2}R=-2\pi \rho_c.
\end{equation}
In the above, we write the energy density $T^{0}_{ \ 0}$ as $-\rho_{\text{NLED}}$, which we may think of as being produced by (\ref{NLEDlagrangian}). Consequently, our single copy gauge field can be understood as being sourced by a combination of the energy density from a non-singular magnetic monopole and the Ricci curvature of the non-singular black hole.

Neither $\rho_{\text{NLED}}$ nor the Ricci scalar dominates the single copy charge density in any region, as can be seen by expanding both quantities in a power series in $r$. For the pure T-dual black hole, the near-origin and asymptotic behaviors of the Ricci scalar are
\begin{equation}
R(r\approx 0)=\frac{24m}{\ell^3}-\frac{90m}{\ell^5}r^2+O(r^4), \ \ \ \ \ R(r\approx\infty)=-\frac{6m\ell^2}{r^5}+O(r^{\text{-}7}),
\end{equation}
while for $\rho_{\text{NLED}}$ stemming from (\ref{NLEDlagrangian}), 
\begin{equation}
\rho_{\text{NLED}}(r\approx 0)=\frac{6m}{g^3}-\frac{15m}{g^5}r^2+O(r^4), \ \ \ \ \ \rho_{\text{NLED}}(r\approx\infty)=\frac{6mg^2}{r^5}-O(r^{\text{-}7}).
\end{equation}
Clearly, $R$ and $\rho_{\text{NLED}}$ have nonzero terms at the same order in $r$. Thus, $\rho_c$ can't `disentangle' contributions from the spacetime curvature from those of the matter energy density.

%========================================================================
%========================================================================
%========================================================================

\subsubsection{Komar energy and net electric charge}

%========================================================================
%========================================================================
%========================================================================

In section \ref{EnergyConditions}, we mentioned that non-singular spacetimes of the type under consideration were studied in the context of the Kerr-Schild double copy in \cite{WiseSphericallySymCDC}. Here, we consider the Komar energy as discussed in \cite{WiseSphericallySymCDC} and verify that sensible results are recovered.  The Komar energy contained in a 3-dimensional (spatial) volume $\Sigma$ is
\begin{equation}\label{Komar1}
E_{K}=2\int_{\Sigma}d^3x\sqrt{\gamma}n_\mu K^\nu R^\mu_{ \ \nu},
\end{equation}
where $K^\nu=(1,0,0,0)$ is the timelike Killing vector, $n^\mu=(1/\sqrt{-g_{tt}},0,0,0)$ is the unit normal to the hypersurface, and $\gamma=r^4\sin^2\theta/g_{tt}$ is the determinant of the induced metric on $\Sigma$. 

Using the relation $R^0_{ \ 0}=\partial^2\phi/2$ from (\ref{RicciAndZeroth}), we can simplify (\ref{Komar1}) to
\begin{equation}
E_{K}=-\int_{\Sigma}r^2drd\Omega\partial^2\phi,
\end{equation}
which illustrates the Komar energy density is
\begin{equation}
\rho_K=-\partial^2\phi=4\pi\rho_c.
\end{equation}
Taking our result\footnote{Note that we have set $\Lambda=0$ for this discussion, as the Komar energy is only defined for asymptotically flat spacetimes. For a somewhat recent discussion on mass in asymptotically de Sitter spacetimes, see \cite{Dolan:2018hpl}.} for $4\pi\rho_c$ from (\ref{rhocharge}), we find after performing the angular integration that 
\begin{equation}
\begin{split}  
E_K&=-4\pi\int_0^{\infty} r^2dr\frac{6m\ell^2(2\ell^2-3r^2)}{(r^2+\ell^2)^{7/2}}\\
&=(4\pi)2m.
\end{split}
\end{equation}
As anticipated, we find the Komar energy is  proportional to the black hole mass. Interestingly, since $\rho_K$ is well defined for $r\in(-\infty,\infty)$, in addition to being symmetric in $r\rightarrow -r, $ the integral can be extended over all $r$, which picks up another factor of 2. On the other hand, if the $r<0$ region is thought of as an alternate universe, then an observer there will also see a Komar energy proportional to the mass in the integral over $(-\infty,0)$. 

Alternatively, from the gauge theory's point of view,
\begin{equation}
4\pi\int_0^{\infty}r^2\rho_c(r)dr\equiv Q_{\text{net}},
\end{equation}
where $Q_{\text{net}}$ is the net electric charge. This relationship between energy in the gravity and charge in the single-copy gauge theory is to be expected, since the  parameter exchange between the gravity side and gauge theory side is 
\begin{equation}\label{masstocharge}
2m\ \ \ \rightarrow \ \ \ Q\sim gT^ac_a
\end{equation}
for the Schwarzschild-to-Coulomb single copy, as was first written down in \cite{MonteiroMainCDC} (here, $g$ is the Yang-Mills coupling and the $T^a$ are the gauge group generators).  Therefore we indeed see a natural association between the Komar energy and the net ``electric'' charge.

%========================================================================
%========================================================================
%========================================================================

\subsubsection{Other non-singular black holes}

%========================================================================
%========================================================================
%========================================================================

\begin{figure}[H]
\centerline{\includegraphics[scale=.4]{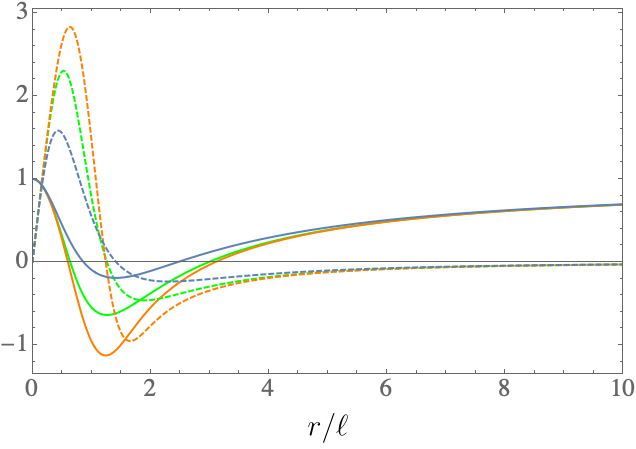}}
\caption{Plot of the metric functions (\ref{otherBHs}) and  electric fields (\ref{otherEs}) for the Hayward (green) and Dymnikova (orange) black holes compared with the T-dual black hole (blue). The electric fields are the dashed curves and the metric functions $-g_{tt}$ are the solid curves.}
\label{otherNSBHplot}
\end{figure}

The essential features of the non-singular single copies are mostly unchanged for the Hayward \cite{Hayward:2005gi} and Dymnikova\footnote{Recently, it was shown in \cite{Ghosh:2020ece} that the Dymnikova solution can also be obtained from a NLED magnetic monopole, with a Lagrangian given by $\mathcal{L}(F)=\frac{3m}{|b|^3}\exp\big[\text{-}\big(2/b^2F\big)^{3/4}\big]$. Their monopole charge $b$ is related to our length parameter by $|b|^3=2m\ell^2$. } \cite{Dymnikova:1992ux} black holes (see also \cite{Carballo-Rubio:2018pmi}), whose metrics are the same form as (\ref{metricg}) but with profile functions
\begin{equation}\label{otherBHs}
-g_{tt}^{\text{H}}(r)=1-\frac{2mr^2}{r^3+2m\ell^2}, \ \ \ \ \ \ \ \ \ -g_{tt}^{\text{D}}(r)=1-\frac{2m}{r}\big(1-e^{-r^3/2m\ell^2}\big).
\end{equation}
Unlike  the T-duality solution, neither of the above are symmetric in $r\rightarrow-r$, and both profiles diverge in the $r<0$ region. However for the $r\in\{0,\infty\}$ domain, the single copies are in essence just scaled versions of the T-duality result. The associated electric fields for each solution are 
\begin{equation}\label{otherEs}
E_r^{\text{H}}(r)=\frac{2mr(4m\ell^2-r^3)}{(r^3+2m\ell^2)^2}, \ \ \ \ \ \ \ \ \ \ \ E_r^{\text{D}}(r)=\frac{2m}{r^2}\Big[e^{-r^3/2m\ell^2}\Big(1+\frac{3r^3}{2m\ell^2}\Big)-1\Big],
\end{equation}
and are shown in Fig. \ref{otherNSBHplot}.

%========================================================================
%========================================================================
%========================================================================

\subsection{Vanishing electric field and horizons}\label{VanEfieldsAndHorizons}

%========================================================================
%========================================================================
%========================================================================

In contrast to the Schwarzschild black hole, the non-singular black holes under consideration are such that $g_{tt}$ has local extrema. The single copy gauge field is $A = (1+g_{tt}(r))dt$, so that the electric field is of the form 
\begin{equation}
    E=\partial_r g_{tt}(r)dr\wedge dt.
\end{equation}
Therefore, the electric field will vanish precisely at the extrema of $g_{tt}.$ To remind the reader, we may write a dimensionless electric field $\tE(\tr)=\ell E(r)$ as in (\ref{dimlessEandrho}),
\begin{equation*}
        \tilde{E}(\tr)=\frac{\alpha\tilde{r}(2-\tilde{r}^2)}{(1+\tilde{r}^2)^{5/2}}+\frac{2\tL}{3}\tr.
\end{equation*}

When $\tL=0,$ we found that the black hole horizon structure was quite sensitive to the critical value of $\alpha_{\text{crit}}=\frac{3\sqrt{3}}{2}.$ In that case, the electric field sees $\alpha$ just as an overall scaling with $\alpha_{\text{crit}}$ being of no special meaning. So in one sense, the information about the horizon on the gravity side is lost through the Kerr-Schild double copy ansatz. We see this in Fig. \ref{rhoandEfigsForMclass}, where each electric field is identical up to overall scaling, while the three different subclasses, $M1,M2$ and $M3$, have radically different horizon structures on the gravity side. 

On the other hand, any static black hole that has at least two horizons is guaranteed to have an extremum between the two radii, say $r_{\pm}$, where $-g_{tt}(r_{\pm})=0$. Hence, we expect a single copy electric field that vanishes at that extremum between $r_{\pm}$. For this reason, the class $M$ and $A$ spacetime's single copy electric field changed sign when approaching $r=0.$ This feature is also exhibited in more standard solutions in general relativity, such as the Reissner-Nordstom black hole and Schwarzschild-de Sitter spacetime. For the nonsingular solutions, in the degenerate cases $M2$, $A2$, and $A4$, the electric field changes signs precisely at the degenerate horizons due to the fact that $-g_{tt}=-\partial_r g_{tt}=0$ at that point.

In the more general case of $\tL\neq 0$, we found that $-g_{tt}$ can also exhibit inflection points (class $B$) or have no extrema aside from the global maximum at $-g_{tt}(r=0)=1$ (class $C$). As a consequence, the single copy electric fields for class $B$ spacetimes vanish precisely at the location of the inflection point \textit{without} changing signs. Electric fields associated with the class $C$ spacetimes retain the same sign respectively in the $r>0$ and $r<0$ regions, only switching signs when passing through the origin. The right panel in Fig. \ref{rhoandEfigsForBCclass} well illustrates this fact.

Summarizing, whenever spacetimes of the form (\ref{metricg}) have two or more horizons, the associated single copy electric field must change signs between the horizons. In the case where $g_{tt}$ has one or zero horizons, the electric field will not switch signs except where $g_{tt}$ possesses extrema. In the exceptional case when $g_{tt}$ has an inflection point, the associated electric field vanishes at that radius without changing signs.

%========================================================================
%========================================================================
%========================================================================

\subsection{Comment on orbits and forces}\label{forcessection}

%========================================================================
%========================================================================
%========================================================================

In this section, we wish to point out a generic feature of static, spherically symmetric metrics and their associated single copy gauge fields. It is well-known that a classical (non-relativistic) particle of charge $q$ in the presence of an external electric field is subject to the Coulomb force $\vec{F}_{c}=q\vec{E}$. Since conservative forces (including the gravitational force) can be written in terms of the gradient of a scalar potential, $\vec{F}_{g}=-\vec{\nabla}V$, we can start from (\ref{efpot}), 
\begin{equation*}
  V_\epsilon(r) =  - \frac{1}{2} g_{tt}\Big( \frac{L^2}{r^2}+\epsilon \Big),
  \end{equation*}
and take the (negative) gradient to obtain the gravitational force on either a massive ($\epsilon=1$) or massless ($\epsilon=0$) particle,
\begin{equation}\label{Fgrav}
F_{\text{grav}}=-\frac{1}{2}E_r(r)\cdot\epsilon-\frac{L^2}{r^2}\Bigg(\frac{g_{tt}(r)}{r}-\frac{1}{2}g_{tt}'(r)\Bigg).
\end{equation}

In the above, we make the identification that $E=\partial_r g_{tt}$, illustrating a very simple but interesting fact: If we imagine a massive particle falling inwards towards the black hole ($L=0$), then 
\begin{equation}
F_{\text{grav}}=-\frac{1}{2}E_r \propto F_{\text{c}},
\end{equation}
where $F_{\text{c}}\sim qE$ is the force on a charge $q$ due to the electric field $E$.  It is of course also true that evaluating $F_{\text{grav}}$ at some black hole horizon $r_H$ is simply the usual expression for surface gravity $\kappa$,
\begin{equation}
\kappa=-\frac{1}{2}\partial_r g_{tt}(r)\Big|_{r_H},
\end{equation}
which is nothing more than the single copy electric field evaluated at the horizon. 

These considerations seem to suggest that in certain special cases, the dynamics of a charged particle in an electromagnetic background produced by the single copy fields are related to the dynamics of a massive particle in the associated gravitational background. This also may hint at a possible connection between quantities in the single copy gauge theory and black hole thermodynamics. 

%We leave further exploration of this notion to follow up work. 

%========================================================================
%========================================================================
%========================================================================

    %%%%%%%%%%%%%%%%%%%%%%%%%%%%%%%%%%%%%%%%%%%%%%%%%%%%%%%%%%%%%%%%%%%%%%%%%%%%%%%%%%%%%%%%
\section{Concluding remarks}\label{conclusion}

%========================================================================
%========================================================================
%========================================================================

In this work, we have explored non-singular black holes and their associated gauge theories in the framework of the classical double copy. Focusing on the T-dual-de Sitter  black hole spacetime, we provided a complete study of the horizon structure and (positive mass) parameter space for the profile function $g_{tt}$,  expanding on the results of \cite{Fernando:2016ksb}. After presenting the Penrose-Carter diagrams for general non-singular black holes, we  studied the energy conditions for the T-dual-de Sitter black hole spacetime, the NLED monopole source of the Bardeen spacetime, before working out a partial treatment of the null and timelike circular orbits for the T-dual black hole.

Following the analysis on the gravitational side of the double copy, we computed the single copy gauge theory quantities $A_\mu, F_{\mu\nu},$ and $\partial_\nu F^{\mu\nu}$, finding each to be non-singular at the origin $r=0.$ These computations  illustrate some simple but interesting relationships between the gauge and gravity sides of the double copy for this spacetime, including a `convention-flip' that associates positive (negative) charge to negative (positive) electric field values in the $r<0$ region. In addition to verifying that expectations for the gauge theory in the interior de Sitter core of the black hole are consistent with \cite{LunaCurvedBckgrnd}, we saw that the Bardeen black hole's single copy charge density $\rho_c$ was a combination of the energy density from the NLED source and the Ricci scalar, however $\rho_c$ could not `see' any difference between the two terms due to each contributing at the same polynomial order in $r$. We next used the T-dual black hole as an example spacetime to confirm the Komar energy maps to the black hole mass which in turn maps to electric charge as outlined in \cite{WiseSphericallySymCDC}, before showing that the gauge theories associated with the Dymnikova and Hayward solutions exhibited qualitatively identical behaviors to that of the T-dual solution in the $r>0$ region.

Possibly our most interesting finding is the interplay between horizons on the gravity side and electric fields on the gauge theory side of the double copy: when the spacetime has multiple horizons, the single copy electric field changes signs between the horizon radii. The reason as to why this happens is straightforward for such simple spacetimes, since the electric field is the first derivative of $g_{tt},$ which must have a local extrema between two horizons.  For all of the extremal solutions, the electric field will have a zero precisely at the merged horizon. In the unique case where the inner, outer, and cosmological horizons all merged (the subclass we called $B2$), we saw that the electric field vanishes yet does not change signs at the location of the triply merged horizon, since $g_{tt}$ has an inflection point at that radius.

There have been very few mentions in the classical double copy literature concerning the role that gravitational horizons play in the single copy gauge theory. Notably, in \cite{TroddenMaxSym}, the authors comment on how the single copy's charge looks qualitatively the same regardless of the black hole mass in the Schwarzschild-de Sitter spacetime (cf. (\ref{masstocharge})), even when the mass takes the maximal value corresponding to the Nariai limit. Here, we have instead found that the single copy electric field's behavior can indicate the presence of a horizon. In order to see this effect, the spacetime must possess at least two horizons for the single copy to `notice' their existence, signaled by $\vec{E}=0.$ 

However, we also showed that $\vec{E}$ can vanish for spacetimes with no horizons at all, as in the traversable wormhole (case $M1$). This behavior arises because $-g_{tt}$ has its local extrema where $-g_{tt}>0$. Therefore, the gauge theory exhibiting $\vec{E}=0$ could be the single copy of a spacetime with multiple horizons or no horizons at all. This degeneracy is  similar to the  fact that taking the square root of a real number always has two solutions corresponding to a positive and negative value. The information of the sign of the solution is then lost when it is squared. This reasoning is consistent with the few additional comments in \cite{TroddenMaxSym} regarding horizons.

Finally, we showed that off of the horizon, the gravitational force on an infalling mass (with zero angular momentum) is  proportional to the force per unit charge on a charged particle in the presence of the single copy electric field. Correspondingly, the electric field evaluated at a horizon maps to the surface gravity (up to a factor of -$\frac{1}{2}$), which may be a hint towards a connection between the gauge theory fields and black hole thermodynamics.

Our findings suggest a number of possible avenues to consider. To start, there are many NLED theories that emit non-singular, static, spherically symmetric solutions. It would be interesting to know which theories double copy to a gravitational theory that also has non-singular solutions, and which do not. In \cite{Pasarin:2020qoa}, for example, it was shown that the non-singular electric field solution to Born-Infeld theory double copies to a gravitational solution that still has curvature singularities at $r=0.$ Caution must be taken, however, since the Kerr-Schild classical double copy maps Maxwell theory onto general relativity. More complex equations of motion on the gauge theory side, as those from a NLED, ought to map to gravitational theories that are more complicated than general relativity. Modifications to the Kerr-Schild single copy ansatz $A_\mu=\phi k_\mu$ might be necessary to make the map between the equations of motion on both sides of the duality consistent.

Spacetimes with multiple horizons are plentiful, and answering whether or not the feature of a vanishing $\vec{E}$ persists between horizons for more complex metrics could illuminate whether the connection between the electric field and horizon structure is generic in the classical double copy. For the Schwarzschild-type ( $g_{rr}=-1/g_{tt}$) static spherically symmetric black hole, we were able to write the associated gauge vector field and corresponding $\vec{E}$ in terms  of an arbitrary $g_{tt}$, thus making a statement about the entire class of such metrics. One issue in extending this analysis to stationary spacetimes is the problem of writing a Kerr-Schild form for an arbitrary metric of that type. It is unclear if the general stationary line element can be written in Kerr-Schild form so that its single copy gauge field is easy to identify. Progress could be made by approaching the problem using the Weyl double copy prescription \cite{LunaTypeD} instead of Kerr-Schild, which uses curved space spinor formalism and does not explicitly rely on the metric being written in Kerr-Schild coordinates. 

We have focused  on spacetimes with positive black hole mass and $\Lambda>0$ and leave a complete study of the $\{\alpha,\tL\}$ parameter space to future work. As discussed in section \ref{sec:nsbh}, allowing for $\Lambda<0$ and/or $m<0$ uncovers regions of the $\{\alpha,\tL\}$ parameter space where the local structure of the spacetime around $r=0$ can be flat or AdS. This is due to the effective cosmological constant being proportional to $\frac{1}{3}\Lambda+\frac{2m}{\ell^3}$. There is also the possibility for the spacetime around $r=0$ to be locally de Sitter, yet tend to AdS as $r\rightarrow\pm\infty$. Because AdS is arguably the most significant spacetime in holography, it may be interesting to investigate the features of the single copy gauge theory in those regions of the parameter space, which may uncover new connections between the double copy and AdS/CFT.

Finally, it would be quite interesting if the physics of black hole thermodynamics has a connection to the classical double copy that is deeper than what we have pointed out in this work. Since the horizon structure of a black hole is intimately connected to its thermodynamics, our observation about the tied behavior of the single-copy electric field and the double-copy gravitational horizon is a possible sign of such a relationship.  Another hint of a relationship is \cite{Keeler:2020rcv} which explored fluid-gravity duality in the context of the classical double copy.  We hope to return to this issue in future work.

%\newpage
%%
\section*{Acknowledgements}
The authors wish to thank Andres Luna and Nikhil Monga for useful conversations while this work was in progress. The work of CK is supported by the U.S. Department of Energy under grant number DE-SC0019470.

%========================================================================
%========================================================================
%========================================================================
\appendix
%========================================================================
%========================================================================
%========================================================================

\section{Curvature tensors and invariants}

The non-vanishing components of the mixed Riemann tensor are:
\bea
R^{tr}{}_{tr} = \frac{m \left(2 \ell^4-11  \ell^2 r^2+2 r^4\right)}{\left(\ell^2+r^2\right)^{7/2}}+\frac{\Lambda }{3} \,, \nonumber \\
R^{t\theta}{}_{t\theta} = R^{t \phi}{}_{t \phi}= R^{r\theta}{}_{r\theta} = R^{r \phi}{}_{r \phi}=\frac{m \left(2  \ell^2-r^2\right)}{\left( \ell^2+r^2\right)^{5/2}}+\frac{\Lambda }{3} \,, \nonumber \\
R^{\theta \phi}{}_{\theta \phi} = \frac{2 m}{\left(\ell^2+r^2\right)^{3/2}}+\frac{\Lambda }{3}\,,
\eea
and those related by symmetries. All of the above approach a finite value in the limit as $|r| \rightarrow 0$: 
\begin{equation}
\frac{2 m}{\ell^3}+\frac{\Lambda }{3} .
\end{equation}
In the large $r$ limit (as $r \rightarrow \infty$), all components of the Riemann tensor approach the asymptotic constant value: $\Lambda/3$. Hence, for all then entire range of the radial coordinate $r\in (-\infty, +\infty)$ the Riemann tensor components are finite.

The non-vanishing components of the mixed Weyl tensor are:
\bea
C^{t\theta}{}_{t\theta} = C^{t \phi}{}_{t \phi}= C^{r\theta}{}_{r\theta} = C^{r \phi}{}_{r \phi}= -\frac{1}{2} C^{tr}{}_{tr}  =  -\frac{1}{2}C^{\theta \phi}{}_{\theta \phi}   \nonumber \\
=\frac{m r^2 \left(3 \ell^2-2 r^2\right)}{2 \left(\ell^2+r^2\right)^{7/2}} \,,
\eea
(and those related by symmetry). Note all the components of the Weyl tensor approach zero in the limit as $r \rightarrow 0$ and in the large $r$ limit, $|r| \rightarrow \infty$.

The non-vanishing components of the mixed Ricci tensor are:
\bea
R^t{}_t = R^r{}_r= \frac{3 \ell^2 m \left(2 \ell^2-3 r^2\right)}{\left(\ell^2+r^2\right)^{7/2}}+\Lambda
\,,
\\
R^\theta{}_\theta = R^\phi{}_\phi= \frac{6 \ell^2 m}{\left(\ell^2+r^2\right)^{5/2}}+\Lambda \,.
\eea
Finally, the non-vanishing components of the mixed Einstein tensor are:
\bea
G^t{}_t = G^r{}_r= -\frac{6 \ell^2 m}{\left(\ell^2+r^2\right)^{5/2}}-\Lambda
\,,
\\
G^\theta{}_\theta = G^\phi{}_\phi= \frac{3 \ell^2 m \left(3 r^2-2 \ell^2\right)}{\left(\ell^2+r^2\right)^{7/2}}-\Lambda \,.
\eea

The Ricci scalar is:
\be\label{rscalar}
R = \frac{ 6 \ell^2 m (4 \ell^2 - r^2)}{(\ell^2 + r^2)^{7/2}} +4 \Lambda 
\ee
The contracted Ricci tensor is:
\be
R_{\mu\nu}R^{\mu\nu} = 2 \left(\frac{6 \ell^2 \Lambda  m \left(4 \ell^2-r^2\right)}{\left(\ell^2+r^2\right)^{7/2}}+\frac{9 \ell^4 m^2 \left(8 \ell^4-4 \ell^2 r^2+13 r^4\right)}{\left(\ell^2+r^2\right)^7}+2 \Lambda ^2\right)
\,,
\ee
which is everywhere finite. 
The Kretschmann scalar is likewise everywhere finite:
\bea
 \mathcal{K}&= &R_{\mu \nu \rho \sigma}R^{\mu \nu \rho \sigma} \nonumber \\
&=&
\frac{12 m^2 \left(8 \ell^8-4 \ell^6 r^2+47 \ell^4 r^4-12 \ell^2 r^6+4 r^8\right)}{\left(r^2 +\ell^2 \right)^7}+ \frac{8 \Lambda  m \left(4 \ell^4-\ell^2 r^2\right)}{\left(r^2 +\ell^2\right)^{7/2}}+\frac{8 \Lambda ^2}{3} \,.
\eea
The contracted Weyl tensor gives:
\be
C_{\mu \nu \rho \sigma}C^{\mu \nu \rho \sigma} =\frac{12 m^2 r^4 \left(3 \ell^2-2 r^2\right)^2}{\left(\ell^2+r^2\right)^7} \,.
\ee

All invariants remain finite for all values of the radial coordinate. Deep inside the black hole the Ricci scalar (\ref{rscalar}) approaches a constant value revealing a de Sitter core:
\be
\lim_{r \to 0} R = \frac{24 m}{\ell^3}+4 \Lambda\,.
\ee
Expanding the metric near $r=0$ gives that of the static de Sitter patch:
\be
g_{tt} \approx - \left(1 - r^2/l^2 + \mathcal{O}(r^4)  \right) \,,
\ee
where we have introduced the interior de Sitter scale 
\be
l^2 =  \frac{3 \ell^3}{\ell^3 \Lambda +6 m} \,.
\ee

%\newpage
\begingroup\raggedright

%\endgroup

%\bibliography{trapbib}
\end{document}